\documentclass[sigconf]{acmart} 

\usepackage{booktabs} 
\usepackage{balance} 

\usepackage{enumitem}
\usepackage{amsmath}
\usepackage{graphicx}
\usepackage{extarrows}
\usepackage{subfigure}
\usepackage{verbatim}
\usepackage{makecell}
\usepackage{diagbox}
\usepackage{bm}
\usepackage{graphics}
\usepackage[export]{adjustbox}
\usepackage{multirow}
\usepackage{float}
\usepackage{placeins}
\usepackage[linesnumbered, ruled]{algorithm2e}
\usepackage{epstopdf}
\usepackage{setspace}
\usepackage[super]{nth}
\usepackage{fancyhdr}
\usepackage{slashbox}
\usepackage{lipsum,multicol}
\usepackage{url}
\usepackage{xcolor}
\usepackage{tikz}

\usepackage[utf8]{inputenc}
\usepackage{array}
\usepackage{makecell}

\def \eg {\emph{e.g.}, }
\def \ie {\emph{i.e.}, }

\newcommand{\model}{ABO\xspace}

\allowdisplaybreaks[4]

\AtBeginDocument{%
  \providecommand\BibTeX{{%
    \normalfont B\kern-0.5em{\scshape i\kern-0.25em b}\kern-0.8em\TeX}}}

\copyrightyear{2025}
\acmYear{2025}
\setcopyright{acmlicensed}
\acmConference[WWW '25]{Proceedings of the ACM Web Conference 2025}{April 28-May 2, 2025}{Sydney, NSW, Australia}
\acmBooktitle{Proceedings of the ACM Web Conference 2025 (WWW '25), April 28-May 2, 2025, Sydney, NSW, Australia}
\acmDOI{10.1145/3696410.3714856}
\acmISBN{979-8-4007-1274-6/25/04}

\begin{document}
 
\title{ABO: Abandon Bayer Filter for Adaptive Edge Offloading in Responsive Augmented Reality} 

\author{Yongxuan Han}
\orcid{0009-0002-3468-2866}
\affiliation{
\institution{Shanghai Jiao Tong University \& \\ State Key Laboratory of Avionics Integration and Aviation System-of-Systems Synthesis}
\city{Shanghai}
\country{China}}
\email{harrisonhan@sjtu.edu.cn}

\author{Shengzhong Liu}
\orcid{0000-0002-7643-7239}
\authornote{Shengzhong Liu is the corresponding author.}
\affiliation{
\institution{Shanghai Jiao Tong University \& \\ State Key Laboratory of Avionics Integration and Aviation System-of-Systems Synthesis}
\city{Shanghai}
\country{China}}
\email{shengzhong@sjtu.edu.cn}

\author{Fan Wu}
\orcid{0000-0003-0965-9058}
\affiliation{
\institution{Shanghai Jiao Tong University}
\city{Shanghai}
\country{China}}
\email{fwu@cs.sjtu.edu.cn}

\author{Guihai Chen}
\orcid{0000-0002-6934-1685}
\affiliation{
\institution{Shanghai Jiao Tong University}
\city{Shanghai}
\country{China}}
\email{gchen@cs.sjtu.edu.cn}

\begin{abstract}
Bayer-patterned color filter array (CFA) has been the go-to solution for color image sensors. 
In augmented reality (AR), although color interpolation (\ie \textit{demosaicing}) of pre-demosaic RAW images facilitates a user-friendly rendering, it creates no benefits in offloaded DNN analytics but increases the image channels by $3\times$ inducing higher transmission overheads. The potential optimization in frame preprocessing of DNN offloading is yet to be investigated.

To that end, we propose \model, an adaptive RAW frame offloading framework that parallelizes demosaicing with DNN computation. 
Its contributions are three-fold:
First, we design a configurable tile-wise RAW image neural codec to compress frame sizes while sustaining downstream DNN accuracy under bandwidth constraints.
Second, based on content-aware tiles-in-frame selection and runtime bandwidth estimation, a dynamic transmission controller adaptively calibrates codec configurations to maximize the DNN accuracy. 
Third, we further optimize the system pipelining to achieve lower end-to-end frame processing latency and higher throughput. 
Through extensive evaluations on a prototype platform, \model consistently achieves 40\% more frame processing throughput and 30\% less end-to-end latency while improving the DNN accuracy by up to 15\% than SOTA baselines. It also exhibits improved robustness against dim lighting and motion blur situations.
\end{abstract}



\maketitle

\renewcommand{\shortauthors}{Yongxuan Han et al.}

\section{Introduction} \label{sec1}

Augmented reality (AR) overlays digital content like images, videos, and sounds onto the real-world environment, enhancing human-environment interactions by adding generated elements.
It has extensive applications in surgical assistance~\cite{shenai2011virtual,chinzei2000mr}, E-commerce~\cite{lu2007augmented,yim2017augmented}, gaming~\cite{ananthanarayanan2017real,piekarski2002arquake}, education~\cite{yuen2011augmented,wu2013current}, and interactive web~\cite{liu2023sa,yang2023tangible}. 
Accurate perception of physical objects in real-time lays the foundational pillars for AR.
While deep neural networks (DNN) have greatly enhanced machine perception capabilities (\eg object detection), their computation-intensive nature incurs significant challenges to limited compute resources on mobile AR devices. 
Thus, edge offloading that transmits data to a nearby edge server for remote DNN analytics has become a leading solution~\cite{eshratifar2018energy,kang2017neurosurgeon}.

The human-device-environment interactions in AR call for low response latency (< 40 ms), high processing throughput (> 25 FPS\cite{24fps}), and sufficient task accuracy for a \textit{responsive}, \textit{smooth}, and \textit{precise} service. 
The end-to-end response latency denotes the duration from a frame being captured to DNN inference results being rendered, including frame preprocessing, data compressing (\ie encoding), two-way transmission, remote DNN inference, and rendering. 
To save network bandwidth, the captured frames are first compressed by an \textit{image codec} before transmission. It should substantially reduce frame sizes, run efficiently on mobile devices, and dynamically adapt to bandwidth fluctuations.
Standard image codecs like JPEG, and recent offloading approaches~\cite{minnen2020channel,mentzer2020high,cheng2020learned}, all focus on encoding demosaiced RGB frames and overlook potential optimizations in preprocessing.
Besides, they are optimized for human browsing experience and could be suboptimal in serving DNN analytics.

Common image sensors use Bayer-patterned color filter array (CFA) to capture color information into single-channel \textit{RAW frames}, which are then \textit{demosaiced} to interpret real-world colors into 3-channel \textit{RGB frames}. 
Demosaicing high-resolution frames can take more than 25 ms on resource-constrained edge devices (\eg NVIDIA Jetson Nano), occupying a large portion of end-to-end latency since the remaining steps only take below 30 ms in total.
We believe demosaicing is only essential to rendering frames into a viewable form but can be skipped for DNN inference.
Since RGB images are solely interpolated from RAW ones without any extra input, RAW images should provide a similar task accuracy when feeding into the downstream DNN model despite having less data.
We are thereby interested in utilizing RAW frames in edge offloading for DNN analytics.
It enables decoupling offloading and demosaicing with no resource contentions (\ie communication vs. computation), thus the onboard demosaicing overhead can be hidden in parallel pipelines, leading to lower latency and higher throughput.

However, there are three technical challenges in the decoupled pipelines:
First, an efficient codec is required to compress RAW frames no larger than JPEG files, while ensuring higher downstream DNN accuracy upon decoding.
Second, to prioritize real-time interactions, we need to dynamically calibrate the configuration of the codec with accuracy-efficiency tradeoffs in response to runtime bandwidth fluctuations.
Third, we need to make both the encoder and the adaptation controller lightweight enough to execute efficiently on edge devices without incurring excessive overhead.

To that end, we propose \model, an adaptive pre-demosaic RAW frame edge offloading framework for DNN analytics, that decouples demosaicing from the offloading into parallel processes, providing real-time responses under constrained and dynamic network bandwidth. 
Its design includes three key perspectives.
First, we train a configurable RAW image neural codec that operates on sub-frame image tiles with an asymmetric autoencoder (\ie device-side shallow encoder and server-side deep decoder) to quickly compress RAW tiles and utilize server-side computation in exchange for transmission savings.
It is trained in a task-aware manner by minimizing both the image reconstruction loss and knowledge distillation loss from downstream DNN, and it is highly configurable in tiles-in-frame selections and the encoded feature dimensions. 
Second, we design a dynamic transmission controller that adaptively decides the two configuration knobs based on real-time video content and estimated bandwidth, to maximize the downstream task accuracy without violating the real-time latency constraint. 
Third, we refactor the system pipeline by decoupling and parallelizing demosaicing and offloading processes to optimize bottleneck resource utilizations (\ie device-side GPU and wireless network).

\begin{figure}[t!]
    \centering
    \includegraphics[width=0.8\linewidth]{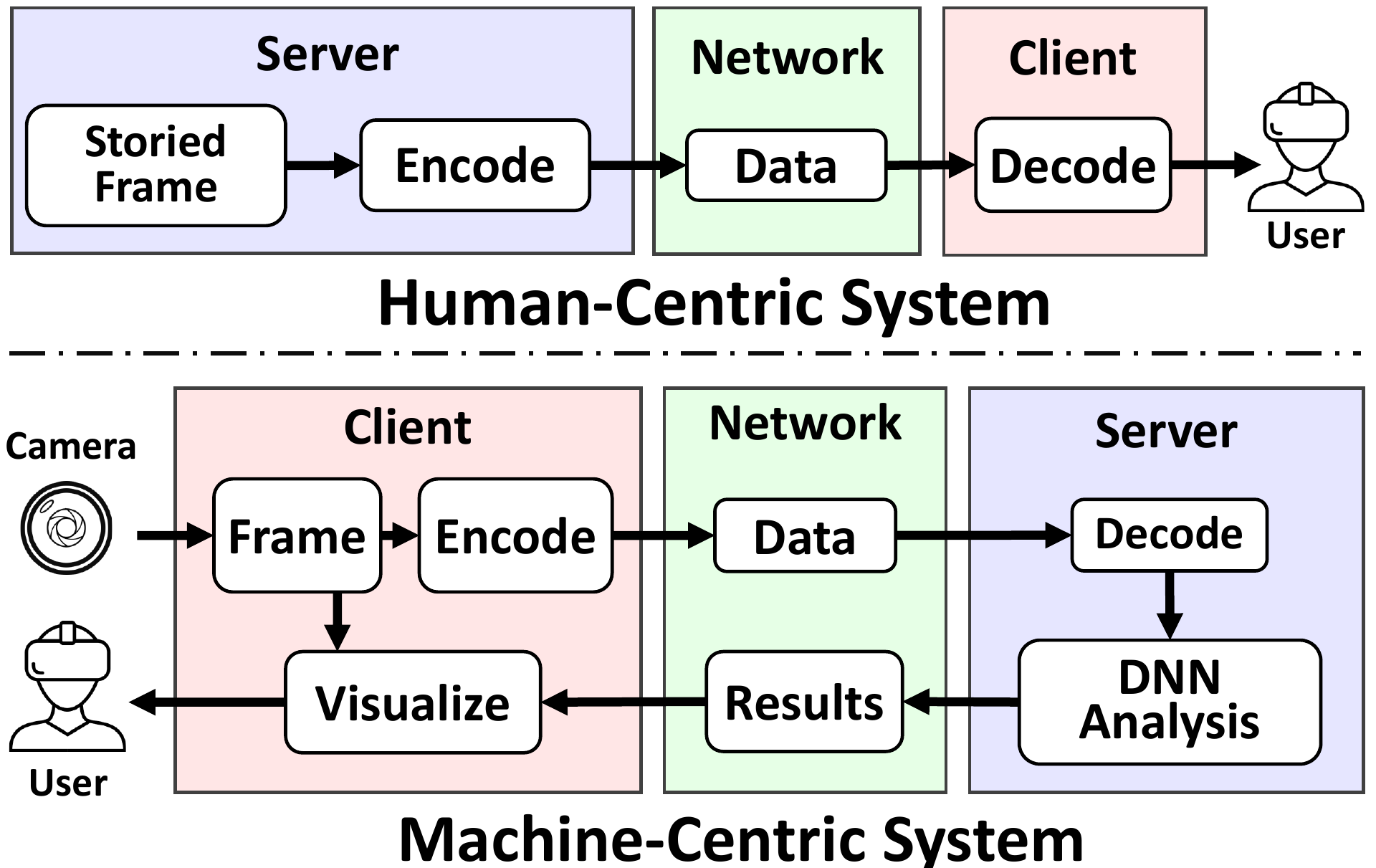}
    \vspace{-0.3cm}
    \caption{Comparing two image transmission types.}
    \label{fig:offload}
\end{figure}

We extensively evaluate the performance of \model with datasets collected with a prototype AR system. The results show that \model outperforms not only JPEG but also SOTA baselines, with up to 15\% improvement on downstream DNN accuracy while incurring similar bandwidth consumption. Meanwhile, the dynamic offloading pipeline of \model consistently provides over 40\% more throughput and 30\% less response latency, being the only one that achieved the target of real-time experience among the tested systems.

Our main contributions are summarized as follows:
\begin{itemize}
    \item We are the first to decouple the visual DNN offloading from image demosaicing such that they can run in parallel for higher throughput and lower end-to-end latency.
    \item We design a tile-based configurable neural codec for RAW images to achieve different latency-accuracy tradeoffs.
    \item We propose an adaptation controller algorithm to optimize the offloaded DNN task accuracy and frame processing throughput upon network bandwidth fluctuations.
    \item We implement both the hardware and software prototype system and perform extensive evaluations to demonstrate the effectiveness and efficiency of \model.
\end{itemize}

\section{Background and Motivations}
\label{sec2}


\subsection{DNN Offloading for Augmented Reality}
DNN analytics (\eg object detection) are crucial in AR applications as they enhance the interactions between digital content and users by helping anchor virtual elements to real-world objects. 
Due to limited onboard resources, offloading computation-intensive DNN tasks to a nearby edge server has been a common practice~\cite{offloadsurvey}: The DNN model loaded on the server processes frames uploaded from the client device. The detection results are then sent back to the client for user-facing rendering.
An offloading pipeline needs to support high-resolution frame processing with high throughput in real time for a responsive, smooth, and accurate user experience.

\begin{figure}[t!]
    \centering
    \includegraphics[width=1.0\linewidth]{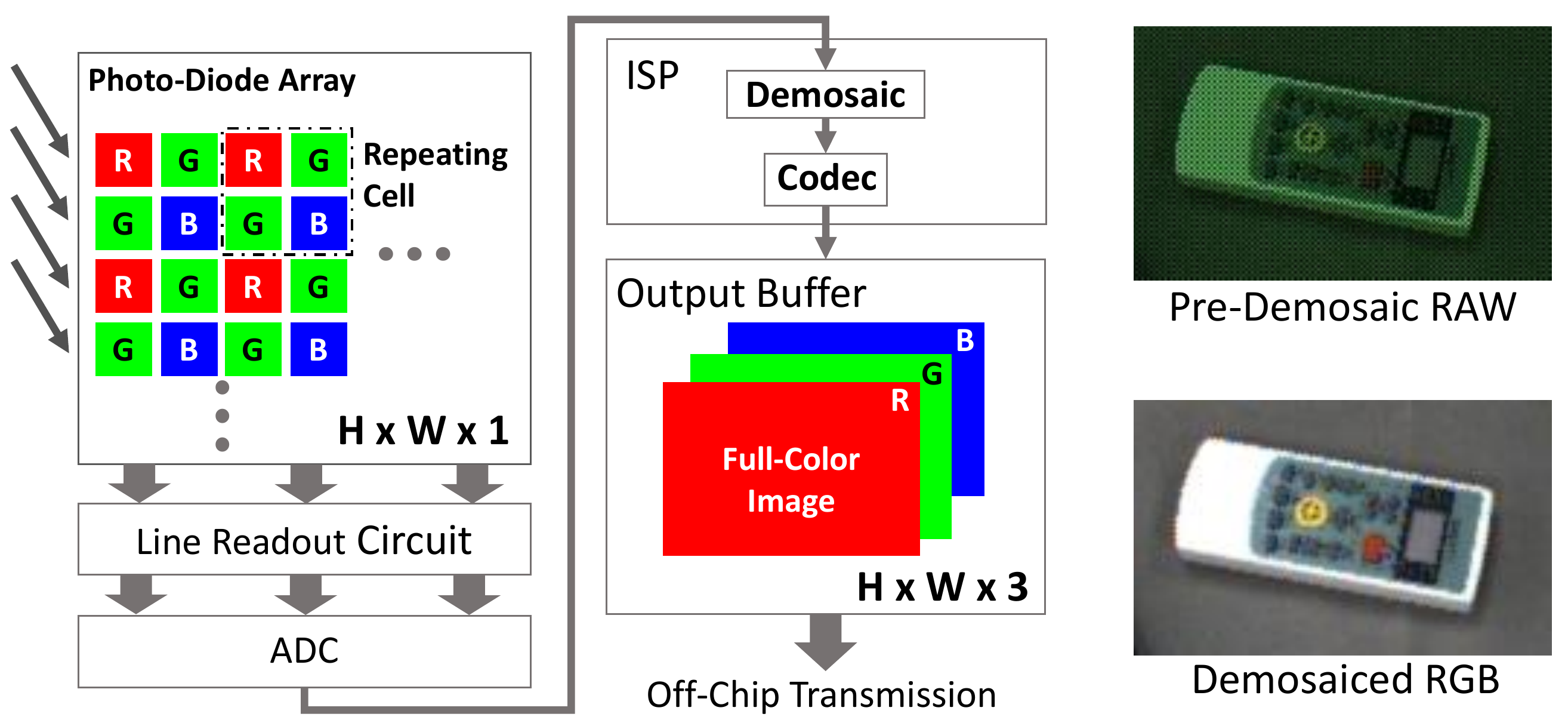}
    \caption{Typical on-sensor image processing pipeline.}
    \label{fig:demosaic}
\end{figure}

To save bandwidth consumption, standard image codecs (\eg JPEG) are applied on the client to compress frames before transmission, which actually mismatch with DNN inference needs.
As shown in Figure~\ref{fig:offload}, we distinguish two transmission types:
\begin{itemize}
    \item In \textbf{human-centric transmission} like content delivery networks, frames are transmitted for human viewing after decoding with standard codecs optimized for human viewing.
    \item In \textbf{machine-centric transmission}, the decoded frames are processed by DNN models, thus downstream model accuracy should guide the codec designing.
\end{itemize}
As a solution, the image codec should be refactored to align with the downstream DNN inference and abandon any information unrelated to DNN prediction (\ie interpolated colors, content intactness). 
Besides, the human-in-the-loop nature of AR calls for both \textit{low latency} and \textit{high throughput}. 
Low latency provides responsiveness, while high throughput guarantees smoothness.

\subsection{Demosaicing in Digital Image Processing}
\label{sec2-1}

Bayer filter is a CFA placed over the photodiode array for capturing images with color. As shown in Figure~\ref{fig:demosaic}, a typical CFA pattern is a 2x2 repeating unit (RGGB filter cell), and it outputs single-channel \textit{RAW images}. 
To reconstruct them into the RGB form, demosaicing algorithms are applied~\cite{demosaic-0,demosaic-1,demosaic-2} to estimate the missing color channels based on nearby pixels, which takes 25 to 45 ms on embedded platforms (\ie Nvidia Jetson Nano, software-based solution, details in Appendix\ref{apdx:softvshard}).
Although demosaiced 3-channel RGB images facilitate human browsing, they do not lead to better DNN inference performance but result in higher transmission overhead. 
Thus the RAW images, with appropriate codecs, could be better candidates for DNN offloading.

\begin{table}[t!]
    \centering
    \caption{Accuracy-bandwidth tradeoffs on object detection between different image codecs. YOLOv5 model is used.}
    \label{tab:t1}
    \vspace{-0.2cm}
    \tabcolsep=0.6cm
    \resizebox{\linewidth}{!}{%
    \begin{tabular}{c|ccc}
    \toprule
    Image Codec & F1 Score    & mAP   & Frame Size \\ \midrule
    RAW    & 0.893 & 0.911 & 242 KB \\ 
    \model-noDistill & 0.873 & 0.878 & 73 $\pm$ 5 KB \\ 
    \model-Distill & 0.923 & 0.937 &  75 $\pm$ 5 KB   \\ 
    \midrule
    RGB    & 0.888 & 0.904 & 726 KB \\ 
    JPEG   & 0.883 & 0.899 &  230 $\pm$ 30 KB   \\
    \bottomrule
    \end{tabular}%
    }
\end{table}


\begin{figure}[t!]
    \centering
    \vspace{0.3cm}
    \includegraphics[width=\linewidth]{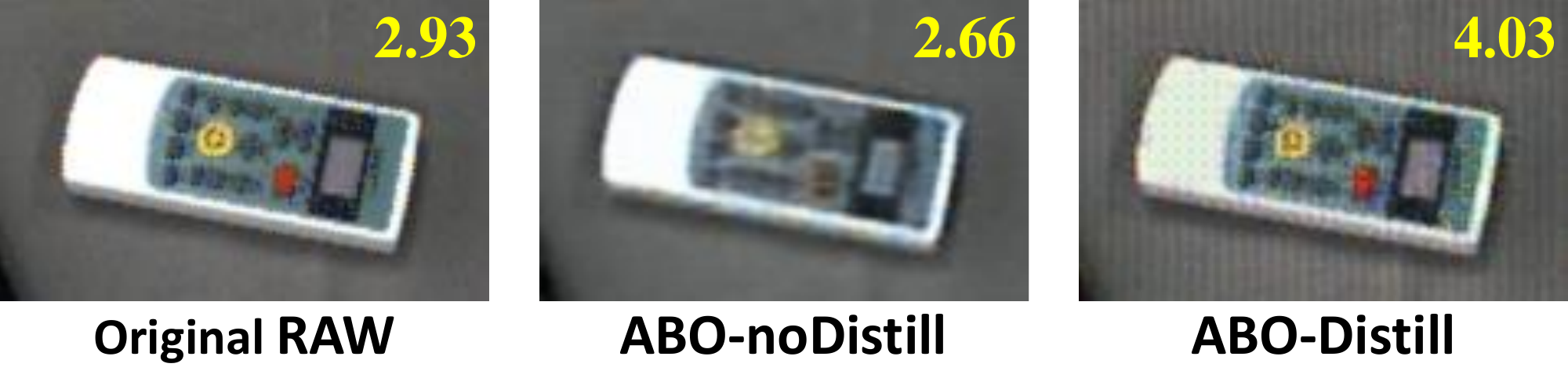}
    \vspace{-0.5cm}
    \caption{Comparison of the original RAW image, \model-noDistill decoded image and \model-Distill decoded image with edge clearness values (Definitions given in Appendix~\ref{apdx:clearness}). All images are demosaiced to RGB formats for better viewing.}
    \label{fig:diff-comparison}
\end{figure}

We conduct experiments to validate the hypothesis. 
Using object detection as the offloaded task, we first train two YOLO models~\cite{yolo} on RAW and RGB images of the same dataset (details in Appendix~\ref{apdx:rawdataset}).
We then evaluate the RGB model on RGB and JPEG-decoded images while evaluating the RAW model on RAW and \model-decoded images, respectively. 
The results in Table~\ref{tab:t1} show that (1) Demosaicing does not enhance the downstream task performance. Instead, the samplings of real-world information remain original and intact in RAW images, which receive better object detection performance than interpolated RGB images (RAW vs. RGB). 
(2) Without encoding, both frame types lead to excessive frame sizes intolerable for transmission, while \model and JPEG can effectively reduce the transmitted frame sizes.
(3) Although \model causes a small degradation in model accuracy (\model-noDistill vs. RAW), finetuning its codec through knowledge distillation achieves even slightly better performance than RAW (\model-Distill vs. RAW).

We visualize the different image types in Figure~\ref{fig:diff-comparison}, and find knowledge distillation achieves targeted image-enhancing artifacts in a mission-oriented manner with better exposed object outlines.
\model achieves lower bandwidth consumption than JPEG without accuracy degradation, thus presenting higher offloading potential.

\subsection{Pre-Demosaic Offloading Savings}
\begin{figure}[t!]
    \centering
    \includegraphics[width=0.9\linewidth]{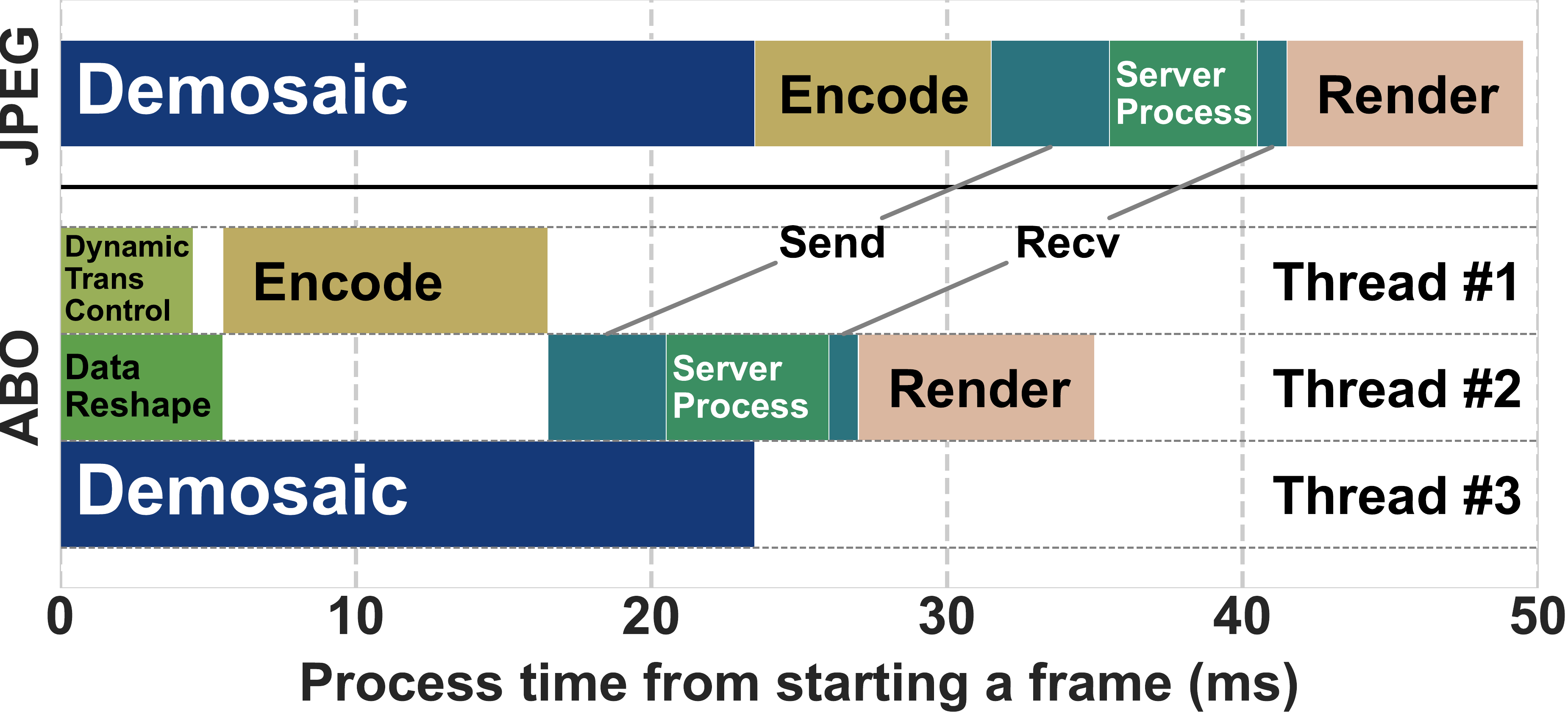}
    \vspace{-0.3cm}
    \caption{Frame processing time comparison between \textit{serialized procedure in JPEG} and \textit{pipelined procedure in \model}.}
    \label{fig:timeline}
    \vspace{0.2cm}
\end{figure}

Existing offloading frameworks~\cite{yao2020deep,li2020reducto,liu2019edge,xie2019source} only optimize the latency from demosaiced RGB images to results rendering but overlook the latency associated with onboard demosaicing.
One main motivation of this paper is to decouple \textit{image offloading} and \textit{demosaicing} for efficiency savings. 
We compare the standard JPEG-based offloading process with the RAW-based ABO process, which allows onboard demosaicing to run in parallel with transmission and remote DNN analysis.
We use an Nvidia Jetson Nano as the edge device and a desktop with Nvidia RTX 4090 GPU as the edge server, whose results are reported in Figure~\ref{fig:timeline}. 
With the decoupled threads, the execution overhead on demosaicing (\ie 25 ms) is hidden behind image offloading steps.
Besides, as indicated in Table~\ref{tab:t1}, single-channel RAW images result in smaller sizes, reducing the network transmission latency.
The end-to-end frame processing latency is reduced from 54 ms to 37 ms, fulfilling the real-time requirement of 40 ms. Its frame throughput increases to over 27 FPS, surpassing the 25 FPS threshold for human viewing.

\section{\model~Framework} 
\label{sec:framework}

\begin{figure}[t!]
    \centering
    \includegraphics[width=1\linewidth]{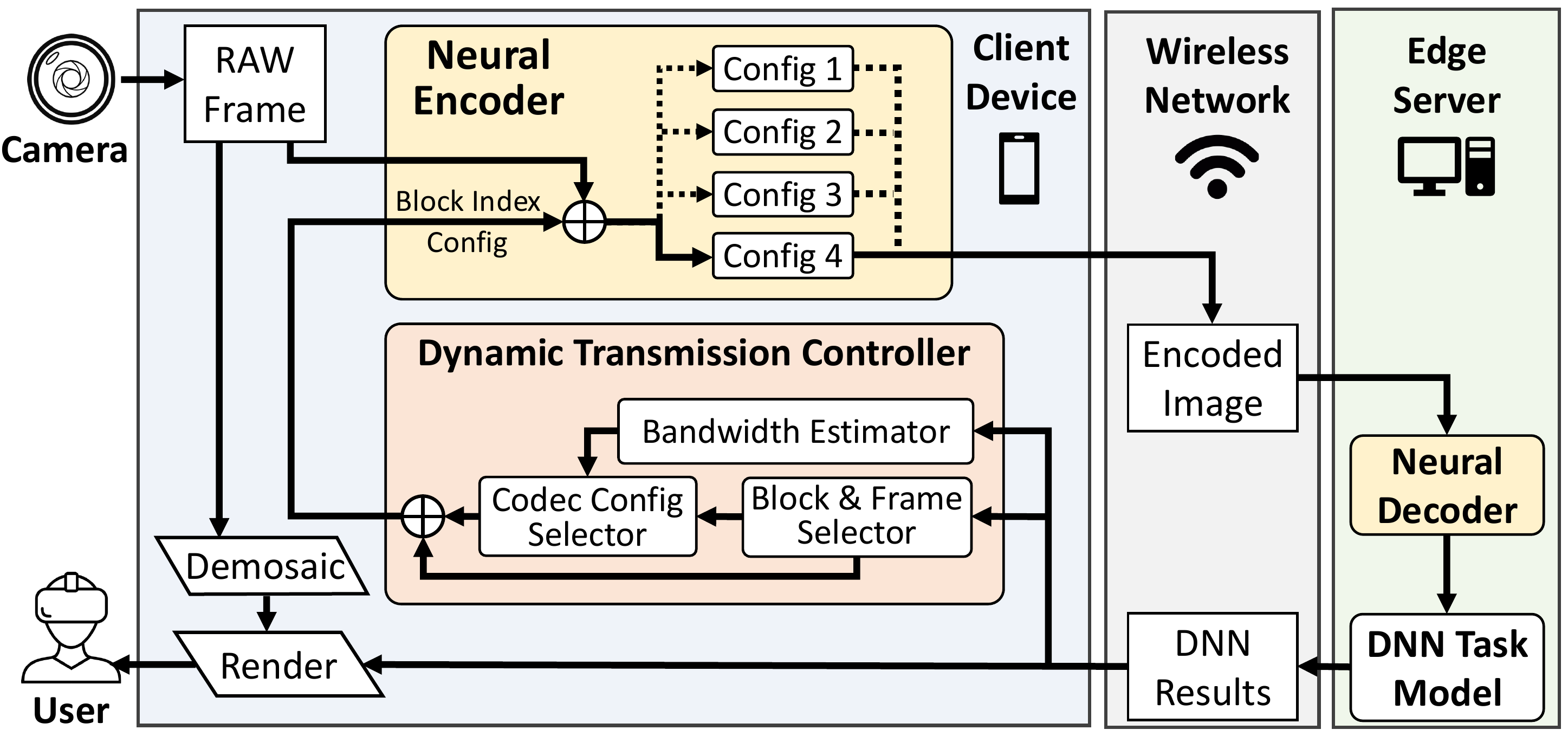}
    \vspace{-0.6cm}
    \caption{\model framework overview.}
    \label{fig:overview}
\end{figure}

\begin{figure*}[t!]
    \centering
    \includegraphics[width=0.85\linewidth]{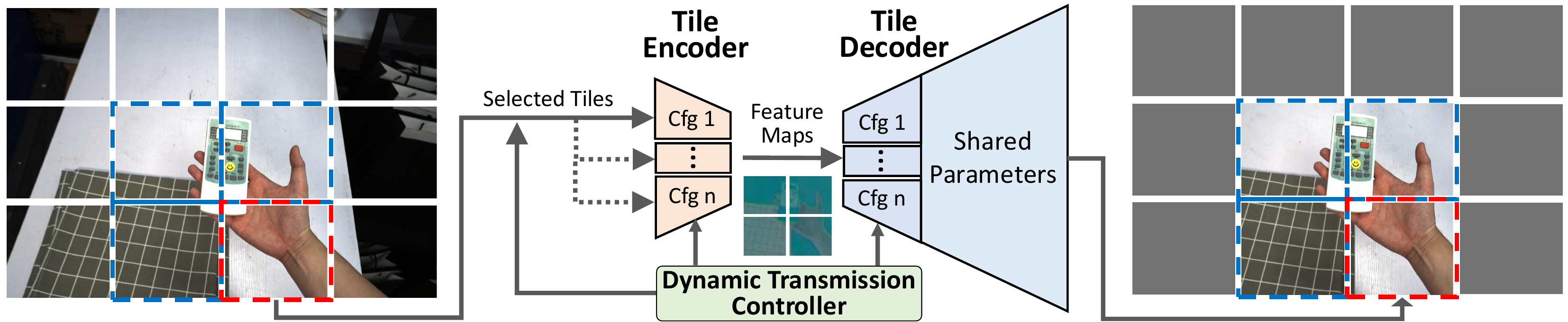}
    \caption{The tile-wise encoding of \model with tile-in-frame selection and multiple encoding configurations.}
    \label{fig:tiles}
\end{figure*}

\subsection{Overview}
The overview of \model is summarized in Figure~\ref{fig:overview}, which includes two main components: a \textit{tile-wise RAW image neural codec} with different configurations to be calibrated with, and a \textit{dynamic transmission controller} tackling bandwidth fluctuations.

\textbf{Tile-wise RAW Image Neural Codec:}
It consists of a lightweight encoder on the edge device to compress sub-frame RAW frame tiles into feature maps with compressed sizes, and a deep decoder on the edge server to reconstruct the RAW image from transmitted feature maps for DNN inference. Multiple encoding configurations are available to balance the offloaded DNN accuracy and bandwidth consumption during the transmission.

\textbf{Dynamic Transmission Controller:}
Given a RAW frame, it first selects the tiles that may contain objects, then determines the codec configuration based on the estimated bandwidth and selected image tiles, adapting to dynamically fluctuating networks.

\subsection{Tile-wise RAW Image Neural Codec}
\label{sec3-2}
A neural codec is expected to significantly reduce the frame size through encoding while effectively sustaining the downstream DNN accuracy on decoded output.
\model's neural codec design is based on two ideas:
First, \model reduces the client-side computation by deploying single-layer neural encoders on edge devices and heavyweight neural decoders on edge servers.
Second, instead of encoding a high-resolution RAW frame as a whole, \model separately encodes individual sub-frame tiles and provides multiple encoder configurations, both constituting a tunable space for codec calibration upon bandwidth dynamics.

\subsubsection{Asymmetric Configurable Autoencoder.}
The neural codec of \model is designed to be asymmetric. 
The encoder only includes a single convolution layer for compressing tile spatial dimensions.
The kernel stride and numbers of output channels are different for each configuration (with parameters separately assigned).
The single convolution layer design can run in real-time on edge devices, while multiple encoders only induce reasonable memory overhead.
On the other hand, The decoder on the edge server (assumed to be resource-rich) has a much larger scale to secure the reconstruction quality.
The basic decoder structure is a stack of residual blocks\cite{he2016deep} (details in Appandix~\ref{apdx:modeulstructure}).
To cope with heterogeneous encoding configurations without hosting multiple deep decoders, as shown in Figure~\ref{fig:tiles}, we use a pluggable design: Each encoder configuration has a separate decoder head to unify the feature map dimensions. After that, all configurations share the same decoder layers until the output.
All encoder-decoder configurations are trained jointly to enhance their compatibility within the shared decoder parameters.

\subsubsection{Frame Color Preservation}
\begin{figure}
    \centering
    \includegraphics[width=0.9\linewidth]{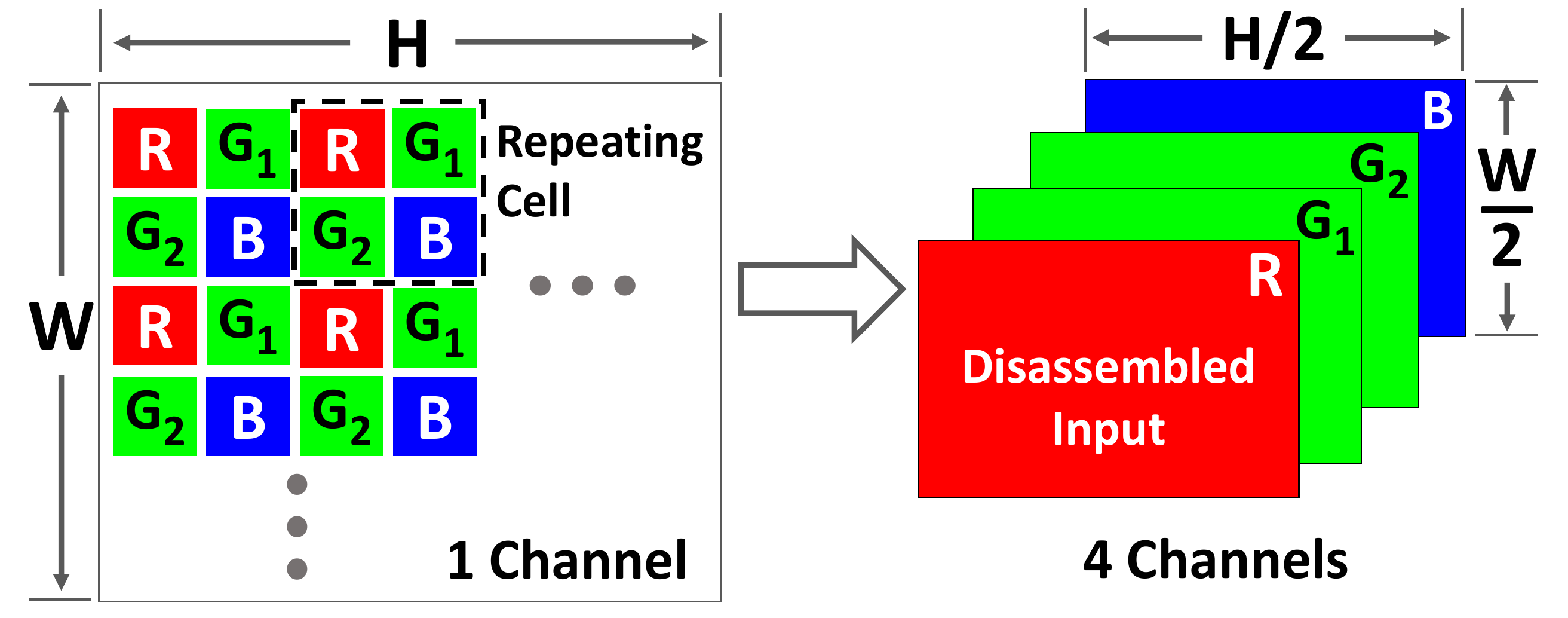}
    \vspace{-0.3cm}
    \caption{Color-preserving input preprocessing.}
    \label{fig:4into1}
\end{figure}

If a RAW frame is fed into the encoder in the original 1-channel format, the grid-patterned RGB information would be erased by the convolution kernel, hard to recover the color information. Instead, we disassemble a RAW frame by the color channels in the CFA repeating pattern and stack the four disassembled channels to create a 4-channel input (illustrated in Figure~\ref{fig:4into1}), so that the mosaiced color information can be preserved during the encoding-decoding processes. 
However, though a disassembled 4-channel and the original 1-channel frames have the same pixel volume and information, the spatial information of a 4-channel input is downsampled by $2\times$ at each dimension which will greatly degrade the extracted high-level spatial features of the DNN model\cite{tan2019efficientnet}.
Thus, the decoded tiles need to be reassembled into the original 1-channel format according to the CFA pattern before feeding into the downstream model.

\subsubsection{Tile Partitioning.}
Object detection models are highly spatially localized in feature extraction, meaning the detection quality will not be affected by removing unrelated background areas. 
Furthermore, selectively transmitting task-related subframe areas helps greatly reduce bandwidth consumption. 
To achieve this, we evenly split a frame as a grid of $r\times c$ tiles with overlap to preserve the reconstruction quality from tile padding artifacts. 
After transmitting the selected tiles with positional indexes to the edge server, the decoded tiles will be placed at their original position in an empty frame canvas filled with default pixel colors. 

\subsubsection{Learning Objectives.}
The neural codec training is divided into two phases.
The first phase only focuses on the reconstruction quality by minimizing a mean-squared error (MSE) loss between the original input and the decoded output. 
In the second phase, we introduce knowledge distillation~\cite{distill} from the downstream DNN model to enhance the preservation of task-related information, achieving higher accuracy in the downstream task.
The frozen downstream model is concatenated to the decoded output and its task loss (\ie object detection loss) is backpropagated to both the encoder and decoder during their update.
Besides, in the distillation phase, losses from different coding configurations are weighted by the proportion of their original MSE losses.
Assume there are $\bm{n}$ codec configurations in total, and denote the MSE loss of one of the configurations as $\bm{\{Loss_{MSE}\}_i}$, the weight for this configuration can be described as:
\begin{equation}
w_i=\frac{\sum_{i=1}^{n}\{Loss_{MSE}\}_i}{\{Loss_{MSE}\}_i}.
\end{equation}
Denote the loss function of downstream DNN visual model as $\bm{Loss_{task}}$, so that the task loss of configuration $i$ is $\bm{\{Loss_{task}\}_i}$, and the knowledge distillation loss is:
\begin{equation}
Loss_{KD}=\sum_{i=1}^{n}w_i*\{Loss_{task}\}_i,
\end{equation}
where $\bm{Loss_{task}}$ depends on the used downstream model. 
In this way, all codec configurations have the same loss value at the beginning of knowledge distillation, with the intuition of forcing them to be updated at a similar rate.

\subsection{Dynamic Transmission Controller}
\label{sec3-3}
To cope with network fluctuations, we design a dynamic transmission controller to continuously calibrate neural codec configurations at runtime. It contains a \textit{codec configuration calibrator} that operates based on an offline profiled look-up table (LUT), a \textit{content-aware tile selector}, and a \textit{lightweight bandwidth estimator}.

\subsubsection{Adaptation Problem Formulation}
The adaptation objective is to maximize offloaded DNN accuracy on each frame satisfying the real-time latency constraint under certain bandwidths. To do so, a LUT is first created offline as $\bm{\{keys:[C_i],values:[(P_i,B_i)]\}_{i=1}^n}$
where $\bm{C_i}$ is the codec configurations, $\bm{P_i}$ and $\bm{B_i}$ are the corresponding profiled task accuracy and bandwidth consumption. $\bm{B_i}$ is calculated by $\bm{B_i=FS_i/T^t_i}$ where $\bm{FS_i}$ and $\bm{T^t_i}$ are the average tile size and a predetermined transmission time threshold. 
If the profiled bandwidth consumption of a configuration is higher than the estimated network bandwidth, the transmission time can exceed the threshold leading to system lagging and stuttering.
At runtime, a content-aware tile-in-frame selection is performed first, giving a list of selected tiles $\bm{LT}$, and the numbers of selected tiles $\bm{t_s}$. Meanwhile, an estimated available bandwidth $\bm{EAB}$ is given by a lightweight estimator. Then the objective of adaptation can be evolved into finding the codec configuration $\bm{C_i}$ with the best task accuracy $\bm{P_i}$ that fits $\bm{B_i<EAB}$.

\subsubsection{Offline Profiling}
To prepare the aforementioned LUT, we use a small profiling data set from the ABO RAW dataset (details in Appendix \ref{apdx:rawdataset}) to measure the average accuracy and tile size. 
The throughput of codec configurations is obtained by running on live cameras under certain bandwidth constraints without involving the dynamic transmission controller.

\subsubsection{Bandwidth Estimation}
We use a lightweight bandwidth estimator to obtain the estimation of the currently available bandwidth (EAB). 
The estimation $\bm{BW}$ is obtained by a simple time differential
$\bm{BW =\frac{S_{trans}}{t_{end}-t_{start}}}$
where $\bm{S_{trans}}$ is the transmitted bit-stream size of the last frame, $\bm{t_{start}}$ and $\bm{t_{end}}$ are the starting and ending timestamp of transmission of the last frame.

\begin{algorithm}[t!]
\SetAlgoNoEnd
\caption{Runtime Configuration Adaptation.}
\label{alg:adaptation}
\KwIn{Offline profiled LUT $LUT=\{keys:[C_i],values:[(P_i,B_i)]\}_{i=1}^n$, 
inference results of last frame $\{bbox_i\}_{i=1}^m$, 
period length $l$, estimated bandwidth $EAB$}
\KwOut{selected tile list $LT$, codec configuration $C_s$}
\SetKwRepeat{Do}{do}{while}
\SetKwFunction{MyFunction}{\text{Generate\_Plan}}

\tcp{Initialization}
Sort $LUT$ by $P_i$ in descending order\;
frame\_count=1\;

\tcp{Main Loop}
\While{TRUE}{
    LT=\{\}\;

    \If{frame\_count==1}{
        $LT=\{T_i\}_{i=1}^{r\times c}$, \  $C_s=C_1$\;
    }
    \Else{
    \For{$i\in\{1,2,\cdots,m\}$ \texttt{and} $t\in\{1,2,\cdots,r\times c\}$}{
        \If{$T_t\cap bbox_i\neq \varnothing$ \texttt{and} $T_t\notin LT$}{
            append $T_t$ into $LT$\;
        }
    }
    num\_tiles = $length(LT)$\;
    \For{$i\in\{1,2,\cdots,n\}$}{
        \If{$B_{i}\times num\_tiles\leq EAB$}{
        $C_s=C_i$\;
        \textbf{break};
        }
    }
}

frame\_count=frame\_count+1 \textbf{if} frame\_count!=l \textbf{else} 1\;
Output $LT$,$C_s$ to encoder\;
$\{bbox_i\}_{i=1}^m$=$\{\hat{bbox}_i\}_{i=1}^k$\; 
}

\end{algorithm}

\subsubsection{Content-Aware Tile Selection}
To enhance content awareness in tile selection, we use detection results from the latest frame as references for tile selection in the next frame.
For each new tile, if it overlaps with any object bounding boxes in its previous frame, it will be selected for encoding and transmission. 
Such a conservative criteria ensures only tiles that are unlikely to contain objects are skipped.
However, if previous detection results are inaccurate, the following tile selections could be affected in a cascade.
To avoid this case, we use a periodic method to reset potential reference errors in fixed-length time windows.
In each window, the first frame is always encoded with all tiles in the highest configuration (called the \textit{key frame}), while the following frames are encoded with tile selection and codec configuration calibration. 
Since key frame transmissions only happen in low frequencies, their overhead is attenuated across all frames in the window and potential reference error accumulation is upper bounded, thus we guarantee the reliability of tile selection with high throughput and low latency even in high-motion scenarios.

\subsubsection{Adaptation Algorithm}
With the above definitions, the adaptation algorithm can be described as shown in Algorithm~\ref{alg:adaptation}.
For a given frame, if it is the first frame of a controlling window (a key frame), all the tiles will be encoded in the highest configuration and transmitted for DNN inference. The inference results will be stored as a reference. If it is not a key frame, the tile selection will be performed first. For each tile in the frame, if it is not already selected and has a cross-section with any bounding box from the last frame inference results, it will be selected for encoding and transmission. After the tile selection is finished, the number of selected tiles will multiplied by the bandwidth usage profile to obtain the estimated bandwidth request (EBR). The best configuration with an EBR less than the EAB will be used for encoding selected tiles and the DNN inference results will be used for tile selection of the next frame.

\begin{figure}[t!]
    \centering
    \includegraphics[width=0.85\linewidth]{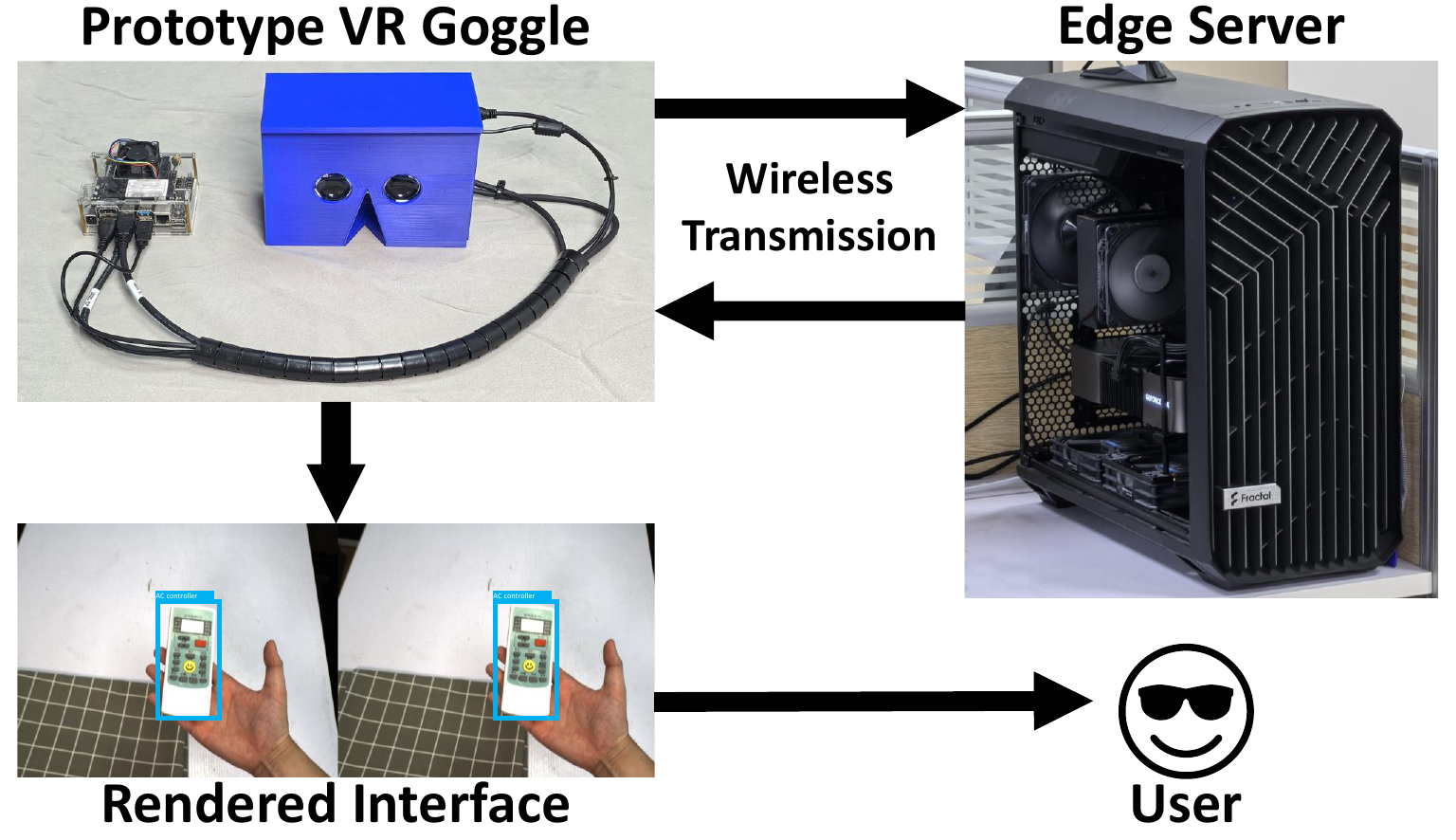}
    \vspace{-0.3cm}
    \caption{3D-printed AR offloading testing platform.}
    \label{fig:prototype}
\end{figure}

\section{Implementation}\label{sec:implementation}
\subsection{Hardware Prototype}
As shown in Figure~\ref{fig:prototype}, we build a hardware prototype system with a 3D-printed AR goggle modified from Google Cardboard~\cite{cardboard}.
We use two IMX178 rolling-shutter sensors from SonySemicon~\cite{hikcamera} as cameras, with a solution of $3072\times2048$ and a maximum frame rate of 60 FPS. 
We use NVIDIA Jetson Nano as the edge device for onboard processing. 
The edge server is configured with an AMD Ryzen R9-7950X CPU with a Nvidia RTX4090 GPU and 64GB RAM, comparable to home desktop PCs. 
The client device and the edge server are connected 
through a 200 Mbps wireless network.

\subsection{Software Implementation}
\label{sec4.2}
We implemented \model framework with 4500 LoC Python code, using PyTorch 2.0~\cite{paszke2019pytorch} as the DNN platform. The network connection and data transmission is achieved through the TCP Socket protocol. The downstream object detection model is YOLOv5-7.0.
Encoders on the edge device are converted to FP16 TensorRT engines~\cite{tensorrt}. Lempel-Ziv coding and Huffman coding are employed for data compressing before transmission to further reduce bandwidth consumption.

\subsection{System Optimizations}
\subsubsection{Pipelining.} 
\begin{figure}[t!]
    \centering
    \includegraphics[width=0.95\linewidth]{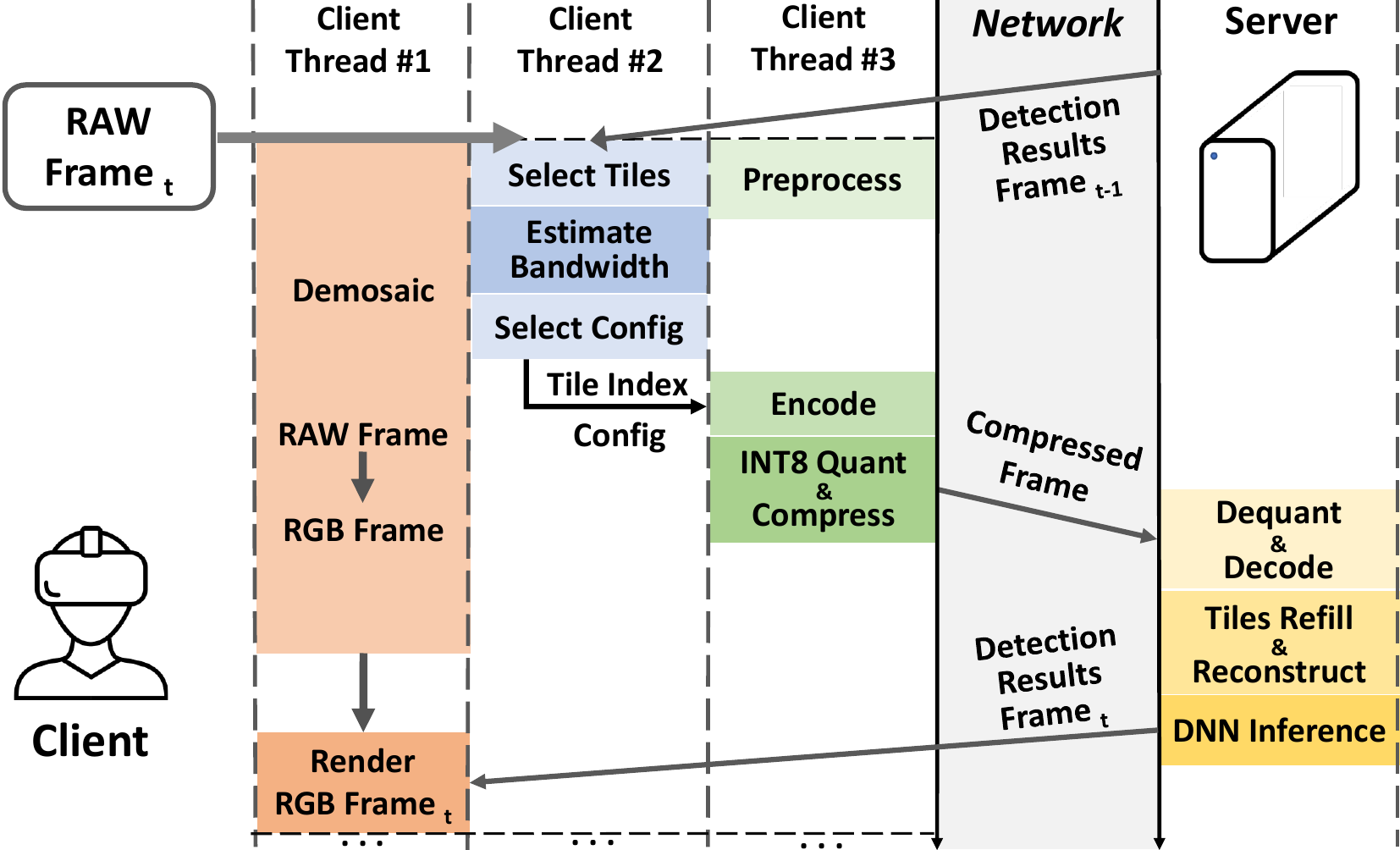}
    \vspace{-0.2cm}
    \caption{\model framework operational workflow.}
    \label{fig:pipeline}
\end{figure}

The pipeline of \model implementation is illustrated in Figure~\ref{fig:pipeline}. Multi-threading the demosaic process with the others means the execution time of all the processes except encoding will be extended since the total CPU resource pool is very limited. However, according to our observation, the multi-threaded pipeline still provides 15\% more FPS with 17\% less end-to-end latency compared to a sequential one.

\subsubsection{Quantization.} 
We reduce bandwidth consumption by quantizing the encoded feature map into unsigned INT8 using simple \textit{add} and \textit{mul} operations. The parameter of quantize is obtained from the upper and lower bound of encoded feature maps of the training set. The quantization brings only about 0.05\% degradation of task performance with a neglectable processing time overhead and a reduction in bandwidth consumption of about 75\%. 

\section{Experiments} \label{sec:experiment}

\subsection{Experimental Setups}
\label{sec:setup}
\subsubsection{Dataset.}
Since existing public datasets do not contain RAW image frames, we manually collect segments of 10-40 seconds consecutive RAW frames at 30 FPS using the prototype AR device, including different possible real-world AR scenarios. The detailed statistics are summarized in Appendix~\ref{apdx:rawdataset}.

\subsubsection{Evaluation metrics}
To verify the performance of the proposed framework, we consider the following metrics perspectives.
\begin{itemize}
    \item \textbf{Task Accuracy:}  
For the object detection task, mean average detection precision (mAP)~\cite{everingham2015pascal, lin2014microsoft}, F1 score, precision, and recall are used to represent the overall task accuracy.
    \item \textbf{Frame Processing Latency:} 
It is defined as the duration from the frame being captured to the remote detection results being rendered and displayed by the client device.
    \item \textbf{Throughput:} 
The frame processing throughput is critical to the smoothness of the user experience. 
It is defined as the number of frames that are processed per second.
    \item \textbf{Bandwidth Consumption:}  
It measures the average size of transmitted frames, as network bandwidth has become one of the resource bottlenecks in edge offloading.
\end{itemize}

\subsection{Compared Baselines}
We compare with the following baselines in our experiments:
\begin{itemize}
    \item \textbf{JPEG~\cite{wallace1992jpeg}:} One of the most widely used methods of lossy compression for RGB images, with a scaled compression ratio reflecting the tradeoff between image quality and file size. We set the JPEG configuration with an adaptive encoding quality module according to the available bandwidth. 
    \item \textbf{DeepCOD~\cite{yao2020deep}:} A neural offloading framework using different layers of deep compression model to achieve efficient transmission while balancing reconstruction quality and downstream task accuracy. 
    \item \textbf{PNC~\cite{wang2023progressive}:} An adaptive neural offloading framework achieving selectable compression level via stochastic tail-drop according to the importance of the feature map layers. 
    \item \textbf{Reducto~\cite{li2020reducto}:} An adaptive frame selection offloading framework with a per-frame differential extractor on the client and offline profiles for scenarios on the server to control which frames to transmit within a stream. 
\end{itemize}

\subsection{Offline Accuracy Profile}

\begin{figure*}[t!]
\centering
\begin{minipage}{.47\linewidth}
  \centering
    \includegraphics[width=1\linewidth]{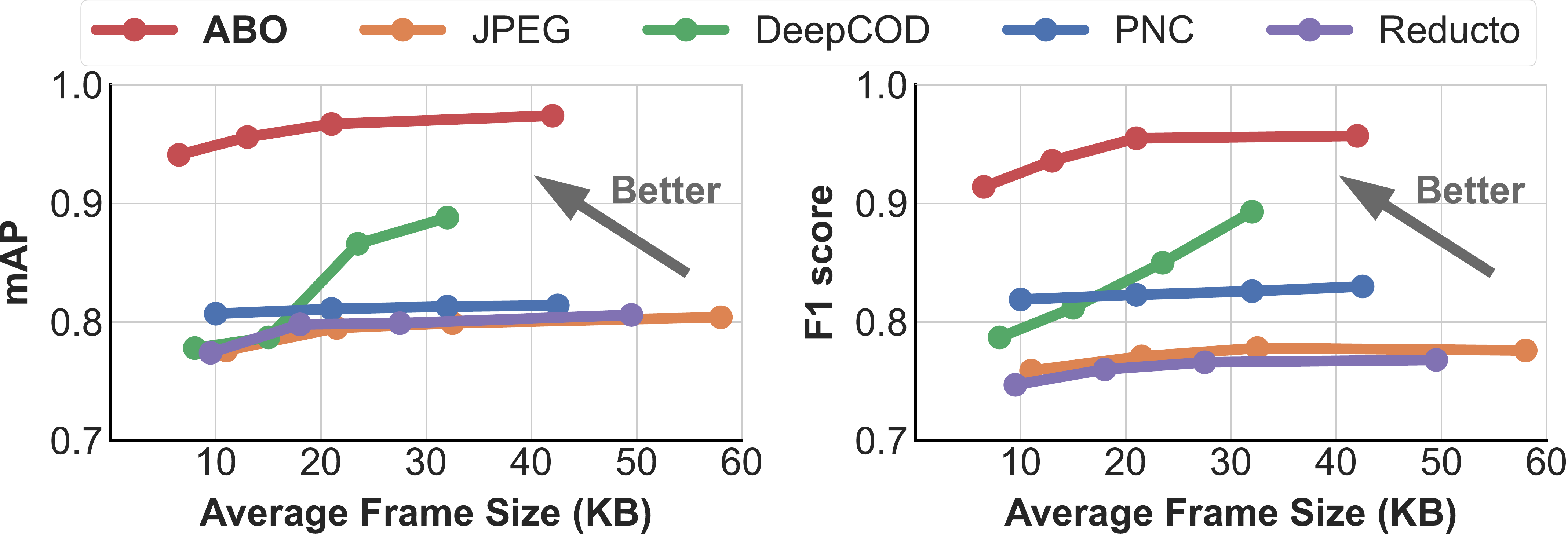}
    \vspace{-0.6cm}
    \caption{Accuracy and frame size profiles.}
    \label{fig:profile acc}
\end{minipage}
\begin{minipage}{0.02\linewidth}

\end{minipage}
\begin{minipage}{.47\linewidth}
  \centering
    \includegraphics[width=1\linewidth]{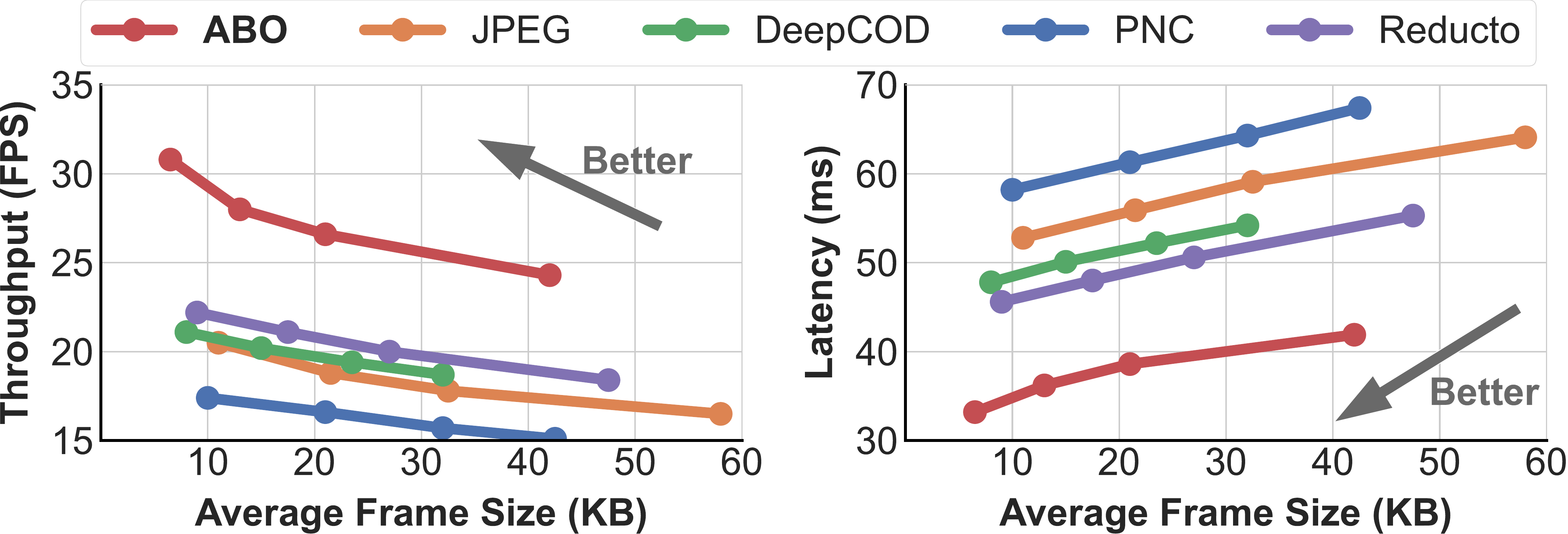}
    \vspace{-0.6cm}
    \caption{Throughput \& latency under 20 Mbps bandwidth.}
    \label{fig:profile lat}
\end{minipage}
\end{figure*}

\begin{figure*}[t!]
    \centering
    \includegraphics[width=0.94\linewidth]{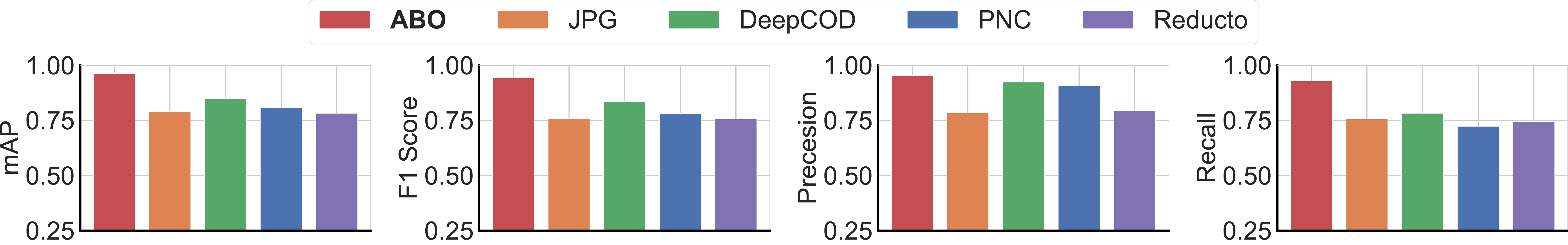}
    \vspace{-0.3cm}
    \caption{Adaptation accuracy under bandwidth dynamics.}
    \label{fig:overall acc}
\end{figure*}

\begin{figure}[t!]
    \centering
    \vspace{0.2cm}
    \includegraphics[width=0.9\linewidth]{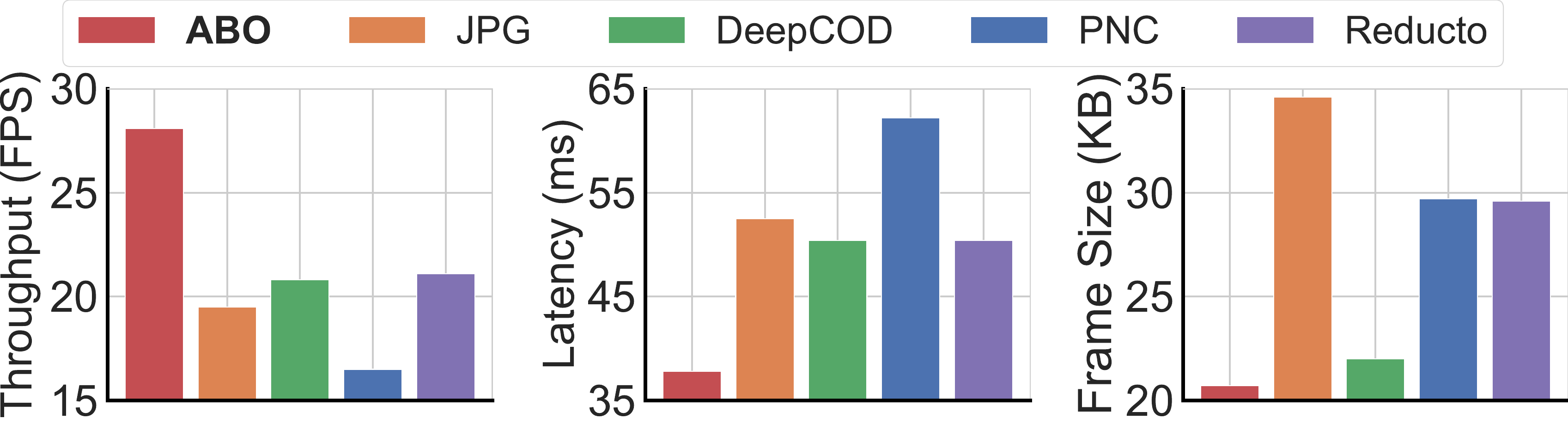}
    \vspace{-0.3cm}
    \caption{Adaptation efficiency under bandwidth dynamics.}
    \label{fig:overall speed}
\end{figure}

An optimal neural codec should reduce the frame into smaller sizes while sustaining higher downstream task accuracy.
We therefore measure the accuracy and bandwidth tradeoffs between different configurations in offline profiling, which represents the Pareto boundaries of the compared frameworks.
Specifically, for each configuration, we measure the task accuracy metrics after decoding and the average compressed frame sizes across the test segments as shown in Figure~\ref{fig:profile acc}.
Each line represents a framework and each point represents a concrete configuration.
\model reaches the highest Pareto limit among all frameworks across all configurations. 
Compared with the JPEG pipeline, \model has an over 15-20\% better task accuracy while only using 30-50\% less bandwidth, demonstrating the importance of task-aware end-to-end code training in DNN offloading.
Besides, although based on the same model architecture, \model outperforms the frame-level encoding of DeepCOD significantly benefiting from its tile-wise encoding and object-free tile filtering during encoding.
Finally, the sole frame selection in Reducto presents poor bandwidth efficiency in downstream tasks.
\subsection{Offline Throughput and Latency Profile}
The end-to-end frame processing latency and the throughput under limited bandwidth are directly related to the responsiveness and smoothness of user experience. 
Therefore, we upper bound the wireless network to 20 Mbps and measure the above two metrics for different configurations within each framework, using all collected test segments.
As illustrated in Figure~\ref{fig:profile lat}, under the same frame sizes, \model achieves the highest throughput and the lowest end-to-end latency, being the only framework surpassing the 25 FPS threshold of real-time inference, demonstrating that the advantages of \model also come from its higher client-side computation efficiency and system pipelining.
Compared to the JPEG pipeline, \model increases the throughput by 50\% while reducing
the end-to-end latency by 35\%, highlighting the savings by parallel demosaicing and DNN offloading. 
PNC struggles with 15\% lower throughput and 10\% higher latency than JPEG, despite its channel-dropping effort, because its encoder still has a large scale that leads to high computation overhead. 
Other baselines have only improved 3\% to 15\% of throughput and 5\% to 10\% of end-to-end latency compared to JPEG without refactoring the underlying codec.

\subsection{Adaptation Performance}

We further perform end-to-end comparisons when individual configurations are integrated into an adaptive framework under network dynamics. 
To conduct the adaptation experiments in a reproducible way, we use both randomly generated and real-world recorded network bandwidth traces~\cite{kan2022improving} (as summarized in Appendix~\ref{apdx:bandtrace}) and replay the network traces to each video segment and each framework, respectively. The network bandwidth is dynamically bounded using Linux Traffic Control (tc)~\cite{linuxtc}.
The results are shown in Figure~\ref{fig:overall acc} and Figure~\ref{fig:overall speed}. \model consistently outperforms the baselines in both downstream task accuracy and frame processing throughput. 
Neural encoding frameworks generally achieve better task performance (\model, DeepCOD, and PNC), where \model demonstrates the best efficiency and accuracy tradeoff. 
DeepCOD and Reducto deliver increased throughput than JPG but still can not meet the smooth browsing requirement (\ie > 25 FPS). 
PNC suffers from long encoding time although the transmitted frame sizes are compressed through its channel-dropping strategy.
In AR applications, since objects are mostly in high motion to the user, the frame selection in Reducto leads to poor downstream accuracy.
Finally, \model~also achieves the lowest end-to-end latency, thanks to its demosaic-free offloading pipeline. 
In summary, the results prove that \model provides the highest quality with guaranteed responsiveness and smoothness to the users.

\subsection{Environmental Robustness}
We evaluate two challenging scenarios in AR applications that require higher robustness (details expanded in Appendix~\ref{apdx:capability}). 

\subsubsection{Low-Light Scenario}
\begin{figure}[t!]
    \centering
    \includegraphics[width=1\linewidth]{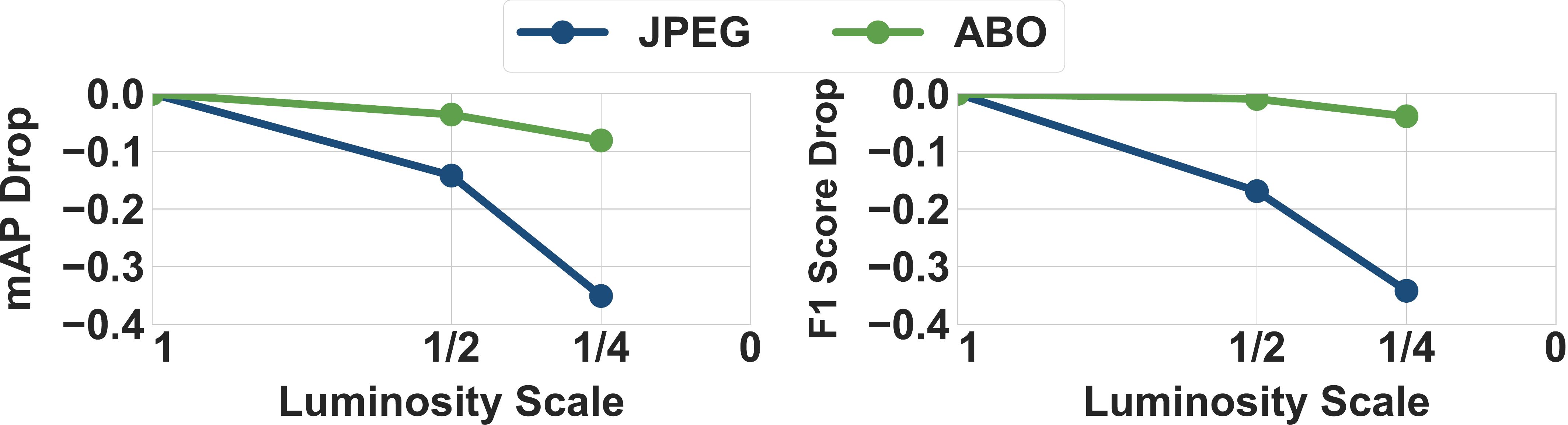}
    \vspace{-0.7cm}
    \caption{Accuracy degradation under low luminosity.}
    \label{fig:luminosity}
\end{figure}

We manually change the luminosity scale of the collected video frames from 1 to 1/2 and 1/4, and compare the accuracy degradation between \model and JPEG codec on their highest configurations.
As shown in Figure~\ref{fig:luminosity}, when operating in low-light conditions, \model achieves lower accuracy degradation below 0.1 while JPEG has an unbearable 0.35 at 1/4 luminosity.

\subsubsection{High-Motion Scenario} 
\begin{figure}[t!]
    \vspace{0.2cm}
    \centering
    \includegraphics[width=1\linewidth]{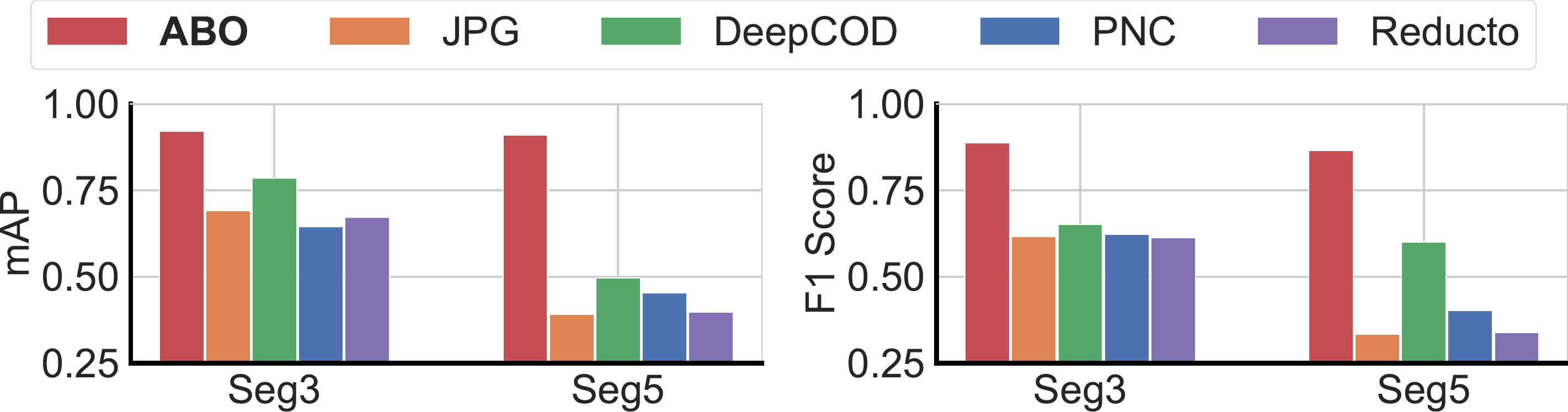}
    \vspace{-0.7cm}
    \caption{Adaptation performance in high-motion videos.}
    \label{fig:motionblur}
\end{figure}

We separately analyze the downstream accuracy of two video segments (\ie Seg3 and Seg5 ) that were collected with higher camera motions.
As shown in Figure~\ref{fig:motionblur}, neural codec solutions tend to have better robustness than frameworks with standard JPEG encoding. 
Besides, \model outperforms the baselines with a larger margin than overall evaluations in Figure~\ref{fig:overall acc}, being the only framework to sustain over 0.85 mAP in both videos.

\subsection{Ablation Study}
To inspect the optimization brought by each module, we designed two ablation experiments. One removes the distillation process in the offline codec training, while the other one removes the tile selection module in dynamic transmission control.
The results are shown in Table~\ref{tab:ablation}. 
Knowledge distillation does not affect frame size, throughput, or end-to-end latency, but brings a huge improvement in task accuracy (over 5\% mAP), making it an essential component in \model. 
Besides, tile selection saves 23\% in average frame sizes with on average 29\% tiles dropped, hence the improvement in throughput and latency, without degrading the downstream accuracy.

\subsection{Overhead Quantification}
\begin{figure}[t!]
    \vspace{0.3cm}
    \centering
    \includegraphics[width=1\linewidth]{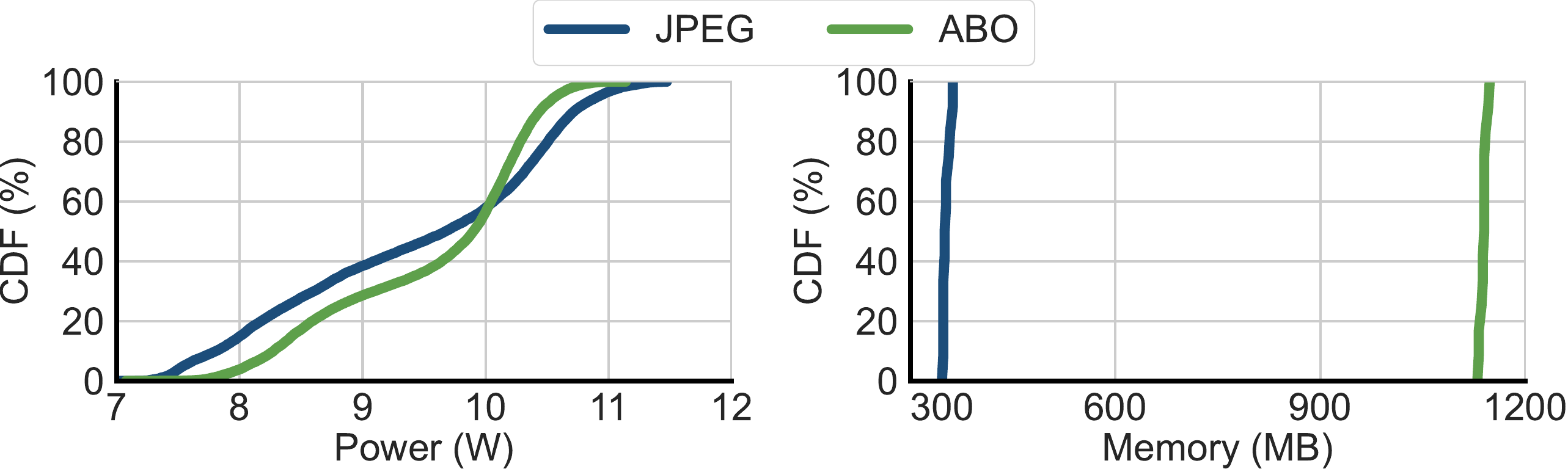}
    \vspace{-0.7cm}
    \caption{Energy and memory overhead quantification.}
    \label{fig:overhead}
\end{figure}

To evaluate the applicability of \model in mobile AR devices,
we monitor the power and memory usage throughout the testing periods. 
The Power consumption is measured with a Monsoon HV power moniter~\cite{hvpm} at 0.2 ms intervals. 
The cumulative overhead distribution curves of JPEG and \model are visualized in Figure~\ref{fig:overhead}.
In energy overhead, the average power of \model is 10 W, which increases by < 10\% than 9.5 W of JPEG, demonstrating its high power efficiency.
In memory usage, after initialization, \model consumes 1080 MB of memory while JPEG consumes 340 MB. 
This is reasonable since we host multiple neural encoders on the edge device, but its absolute memory consumption is acceptable to current AR device capacities.

\section{Related Work} 
\label{sec:6_related}

Despite other creative solutions~\cite{3ccd,x3sensor},
Bayer-patterned CFA has been the go-to solution for color image sensors. Encoding images with DNN models has been rapidly developed in recent years~\cite{toderici2017full,liu2018deepn,toderici2015variable,liu2022self}. Despite that several works have encoded CFA images~\cite{chung2008lossless,lakshmi2016visually,richter2021bayer,abdorgchen2018learning,dong2022abandoning}, there is still a long way to using neural codecs for CFA images on mobile devices with insufficient computing power. Traditional DNN methods targeting to improve
human-view-experience by using more parameters in the DNN codec which induces huge time overhead and performance consumption, making any utilization in mobile computing scenario impractical. Hardware methods have also been attempted~\cite{ma2023leca}, despite the outstanding performance, a custom-made sensor module is too costly for low-end devices. 
Attempts to use Bayer image for object detection have also been made~\cite{lu2023object}, but it requires a distinct model that only targets a single mission. 

The offloading technique is proposed to address the restraint of insufficient computing power of edge devices~\cite{xiao2021task,he2022pyramid,liu2022adamask}. Using such technique in AR/VR scenarios has been a new investigation direction~\cite{kong2023accumo,huang2023re,guan2023metastream,cheng2024grace}. Among existing solutions, there are several frameworks considering the rising trend of using high-resolution to improve the quality of user experience~\cite{zhang2021elf,ren2021adaptive,wang2023real}. However, there is no investigation into using pre-debayered RAW images in mobile offloading scenarios, except \model.

\begin{table}[t!]
    \centering
    \caption{\model ablation study results.}
    \vspace{-0.2cm}
    \label{tab:ablation}
    \resizebox{\linewidth}{!}{%
    \begin{tabular}{c|ccccc}
    \toprule
      & F1 Score & mAP & Avg Frame Size & Throughput & Latency \\ \midrule
    \model & 0.941 & 0.963 & 19.6 KB & 29.0 FPS & 37.1 ms \\ 
    \midrule
    \model-noDistill & 0.875 & 0.907 & 20.4 KB & 28.9 FPS & 37.3 ms \\
    \model-allTiles & 0.942 & 0.963 & 25.4 KB & 27.2 FPS & 39.2 ms \\
    \bottomrule
    \end{tabular}%
    }
\end{table}

\section{Conclusion} 
\label{sec:conclusion}

In this paper, we introduced \model, an adaptive RAW frame offloading framework for DNN analytics in AR applications. 
It relies on three main designs:
First, it decouples demosaicing and offloading into parallel processes through system pipelining for reduced latency;
Second, it contains a tile-wise RAW image neural codec with multiple configurations;
Finally, it adaptively calibrates the coding configurations based on the content-aware tile selection and runtime bandwidth.
Through evaluations on a prototyped hardware platform, \model constantly achieved up to 15\% improvement in downstream task accuracy while increasing the frame processing throughput by 40\% and reducing the end-to-end latency by 30\% with similar bandwidth consumption, compared to SOTA baselines. 

\begin{acks}
The authors would like to thank Maozhe Zhao for assisting in the paper's presentation and general writing.
This work was sponsored in part by the National Key R\&D Program of China (No. 2022ZD0119100), in part by China NSF grant No. 62472278, 62025204, 62432007, 62332014, and 62332013, in part by Alibaba Group through Alibaba Innovation Research Program, and in part by Tencent Rhino Bird Key Research Project. 
This work was partially supported by SJTU Kunpeng \& Ascend Center of Excellence.
The opinions, findings, conclusions, and recommendations expressed in this paper are those of the authors and do not necessarily reflect the views of the funding agencies or the government.
\end{acks}

\newpage
\bibliographystyle{ACM-Reference-Format}
\balance
\bibliography{reference}


\begin{thebibliography}{64}


\ifx \showCODEN    \undefined \def \showCODEN     #1{\unskip}     \fi
\ifx \showISBNx    \undefined \def \showISBNx     #1{\unskip}     \fi
\ifx \showISBNxiii \undefined \def \showISBNxiii  #1{\unskip}     \fi
\ifx \showISSN     \undefined \def \showISSN      #1{\unskip}     \fi
\ifx \showLCCN     \undefined \def \showLCCN      #1{\unskip}     \fi
\ifx \shownote     \undefined \def \shownote      #1{#1}          \fi
\ifx \showarticletitle \undefined \def \showarticletitle #1{#1}   \fi
\ifx \showURL      \undefined \def \showURL       {\relax}        \fi
\providecommand\bibfield[2]{#2}
\providecommand\bibinfo[2]{#2}
\providecommand\natexlab[1]{#1}
\providecommand\showeprint[2][]{arXiv:#2}

\bibitem[car(2014)]%
        {cardboard}
 \bibinfo{year}{2014}\natexlab{}.
\newblock \bibinfo{booktitle}{\emph{Google Cardboard}}.
\newblock
\urldef\tempurl%
\url{https://arvr.google.com/cardboard}
\showURL{%
\tempurl}


\bibitem[lin(2019)]%
        {linuxtc}
 \bibinfo{year}{2019}\natexlab{}.
\newblock \bibinfo{booktitle}{\emph{Ubuntu tc8}}.
\newblock
\urldef\tempurl%
\url{https://manpages.ubuntu.com/manpages/focal/man8/tc.8.html}
\showURL{%
\tempurl}


\bibitem[ten(2024)]%
        {tensorrt}
 \bibinfo{year}{2024}\natexlab{}.
\newblock \bibinfo{booktitle}{\emph{NVIDIA TensorRT}}.
\newblock
\urldef\tempurl%
\url{https://developer.nvidia.com/tensorrt}
\showURL{%
\tempurl}


\bibitem[Ananthanarayanan et~al\mbox{.}(2017)]%
        {ananthanarayanan2017real}
\bibfield{author}{\bibinfo{person}{Ganesh Ananthanarayanan}, \bibinfo{person}{Paramvir Bahl}, \bibinfo{person}{Peter Bod{\'\i}k}, \bibinfo{person}{Krishna Chintalapudi}, \bibinfo{person}{Matthai Philipose}, \bibinfo{person}{Lenin Ravindranath}, {and} \bibinfo{person}{Sudipta Sinha}.} \bibinfo{year}{2017}\natexlab{}.
\newblock \showarticletitle{Real-time video analytics: The killer app for edge computing}.
\newblock \bibinfo{journal}{\emph{computer}} \bibinfo{volume}{50}, \bibinfo{number}{10} (\bibinfo{year}{2017}), \bibinfo{pages}{58--67}.
\newblock


\bibitem[Brownlow(1980)]%
        {24fps}
\bibfield{author}{\bibinfo{person}{Kevin Brownlow}.} \bibinfo{year}{1980}\natexlab{}.
\newblock \bibinfo{booktitle}{\emph{What Was the Right Speed?}}
\newblock
\urldef\tempurl%
\url{https://cinemaweb.com/silentfilm/bookshelf/18_kb_2.htm}
\showURL{%
\tempurl}


\bibitem[Chen et~al\mbox{.}(2018)]%
        {abdorgchen2018learning}
\bibfield{author}{\bibinfo{person}{Chen Chen}, \bibinfo{person}{Qifeng Chen}, \bibinfo{person}{Jia Xu}, {and} \bibinfo{person}{Vladlen Koltun}.} \bibinfo{year}{2018}\natexlab{}.
\newblock \showarticletitle{Learning to see in the dark}. In \bibinfo{booktitle}{\emph{Proceedings of the IEEE conference on computer vision and pattern recognition}}. \bibinfo{pages}{3291--3300}.
\newblock


\bibitem[Cheng et~al\mbox{.}(2024)]%
        {cheng2024grace}
\bibfield{author}{\bibinfo{person}{Yihua Cheng}, \bibinfo{person}{Ziyi Zhang}, \bibinfo{person}{Hanchen Li}, \bibinfo{person}{Anton Arapin}, \bibinfo{person}{Yue Zhang}, \bibinfo{person}{Qizheng Zhang}, \bibinfo{person}{Yuhan Liu}, \bibinfo{person}{Kuntai Du}, \bibinfo{person}{Xu Zhang}, \bibinfo{person}{Francis~Y Yan}, {et~al\mbox{.}}} \bibinfo{year}{2024}\natexlab{}.
\newblock \showarticletitle{GRACE: Loss-Resilient Real-Time Video through Neural Codecs}. In \bibinfo{booktitle}{\emph{21st USENIX Symposium on Networked Systems Design and Implementation (NSDI 24)}}. \bibinfo{pages}{509--531}.
\newblock


\bibitem[Cheng et~al\mbox{.}(2020)]%
        {cheng2020learned}
\bibfield{author}{\bibinfo{person}{Zhengxue Cheng}, \bibinfo{person}{Heming Sun}, \bibinfo{person}{Masaru Takeuchi}, {and} \bibinfo{person}{Jiro Katto}.} \bibinfo{year}{2020}\natexlab{}.
\newblock \showarticletitle{Learned image compression with discretized gaussian mixture likelihoods and attention modules}. In \bibinfo{booktitle}{\emph{Proceedings of the IEEE/CVF conference on computer vision and pattern recognition}}. \bibinfo{pages}{7939--7948}.
\newblock


\bibitem[Chinzei et~al\mbox{.}(2000)]%
        {chinzei2000mr}
\bibfield{author}{\bibinfo{person}{Kiyoyuki Chinzei}, \bibinfo{person}{Nobuhiko Hata}, \bibinfo{person}{Ferenc~A Jolesz}, {and} \bibinfo{person}{Ron Kikinis}.} \bibinfo{year}{2000}\natexlab{}.
\newblock \showarticletitle{MR compatible surgical assist robot: System integration and preliminary feasibility study}. In \bibinfo{booktitle}{\emph{Medical Image Computing and Computer-Assisted Intervention--MICCAI 2000: Third International Conference, Pittsburgh, PA, USA, October 11-14, 2000. Proceedings 3}}. Springer, \bibinfo{pages}{921--930}.
\newblock


\bibitem[Chung and Chan(2008)]%
        {chung2008lossless}
\bibfield{author}{\bibinfo{person}{King-Hong Chung} {and} \bibinfo{person}{Yuk-Hee Chan}.} \bibinfo{year}{2008}\natexlab{}.
\newblock \showarticletitle{A lossless compression scheme for Bayer color filter array images}.
\newblock \bibinfo{journal}{\emph{IEEE Transactions on Image Processing}} \bibinfo{volume}{17}, \bibinfo{number}{2} (\bibinfo{year}{2008}), \bibinfo{pages}{134--144}.
\newblock


\bibitem[Deng et~al\mbox{.}(2009)]%
        {deng2009imagenet}
\bibfield{author}{\bibinfo{person}{Jia Deng}, \bibinfo{person}{Wei Dong}, \bibinfo{person}{Richard Socher}, \bibinfo{person}{Li-Jia Li}, \bibinfo{person}{Kai Li}, {and} \bibinfo{person}{Li Fei-Fei}.} \bibinfo{year}{2009}\natexlab{}.
\newblock \showarticletitle{Imagenet: A large-scale hierarchical image database}. In \bibinfo{booktitle}{\emph{2009 IEEE conference on computer vision and pattern recognition}}. Ieee, \bibinfo{pages}{248--255}.
\newblock


\bibitem[Dong et~al\mbox{.}(2022)]%
        {dong2022abandoning}
\bibfield{author}{\bibinfo{person}{Xingbo Dong}, \bibinfo{person}{Wanyan Xu}, \bibinfo{person}{Zhihui Miao}, \bibinfo{person}{Lan Ma}, \bibinfo{person}{Chao Zhang}, \bibinfo{person}{Jiewen Yang}, \bibinfo{person}{Zhe Jin}, \bibinfo{person}{Andrew Beng~Jin Teoh}, {and} \bibinfo{person}{Jiajun Shen}.} \bibinfo{year}{2022}\natexlab{}.
\newblock \showarticletitle{Abandoning the bayer-filter to see in the dark}. In \bibinfo{booktitle}{\emph{Proceedings of the ieee/cvf conference on computer vision and pattern recognition}}. \bibinfo{pages}{17431--17440}.
\newblock


\bibitem[Eshratifar and Pedram(2018)]%
        {eshratifar2018energy}
\bibfield{author}{\bibinfo{person}{Amir~Erfan Eshratifar} {and} \bibinfo{person}{Massoud Pedram}.} \bibinfo{year}{2018}\natexlab{}.
\newblock \showarticletitle{Energy and performance efficient computation offloading for deep neural networks in a mobile cloud computing environment}. In \bibinfo{booktitle}{\emph{Proceedings of the 2018 on Great Lakes Symposium on VLSI}}. \bibinfo{pages}{111--116}.
\newblock


\bibitem[Everingham et~al\mbox{.}(2015)]%
        {everingham2015pascal}
\bibfield{author}{\bibinfo{person}{Mark Everingham}, \bibinfo{person}{SM~Ali Eslami}, \bibinfo{person}{Luc Van~Gool}, \bibinfo{person}{Christopher~KI Williams}, \bibinfo{person}{John Winn}, {and} \bibinfo{person}{Andrew Zisserman}.} \bibinfo{year}{2015}\natexlab{}.
\newblock \showarticletitle{The pascal visual object classes challenge: A retrospective}.
\newblock \bibinfo{journal}{\emph{International journal of computer vision}}  \bibinfo{volume}{111} (\bibinfo{year}{2015}), \bibinfo{pages}{98--136}.
\newblock


\bibitem[Guan et~al\mbox{.}(2023)]%
        {guan2023metastream}
\bibfield{author}{\bibinfo{person}{Yongjie Guan}, \bibinfo{person}{Xueyu Hou}, \bibinfo{person}{Nan Wu}, \bibinfo{person}{Bo Han}, {and} \bibinfo{person}{Tao Han}.} \bibinfo{year}{2023}\natexlab{}.
\newblock \showarticletitle{Metastream: Live volumetric content capture, creation, delivery, and rendering in real time}. In \bibinfo{booktitle}{\emph{Proceedings of the 29th Annual International Conference on Mobile Computing and Networking}}. \bibinfo{pages}{1--15}.
\newblock


\bibitem[He et~al\mbox{.}(2016)]%
        {he2016deep}
\bibfield{author}{\bibinfo{person}{Kaiming He}, \bibinfo{person}{Xiangyu Zhang}, \bibinfo{person}{Shaoqing Ren}, {and} \bibinfo{person}{Jian Sun}.} \bibinfo{year}{2016}\natexlab{}.
\newblock \showarticletitle{Deep residual learning for image recognition}. In \bibinfo{booktitle}{\emph{Proceedings of the IEEE conference on computer vision and pattern recognition}}. \bibinfo{pages}{770--778}.
\newblock


\bibitem[He et~al\mbox{.}(2022)]%
        {he2022pyramid}
\bibfield{author}{\bibinfo{person}{Qiang He}, \bibinfo{person}{Zeqian Dong}, \bibinfo{person}{Feifei Chen}, \bibinfo{person}{Shuiguang Deng}, \bibinfo{person}{Weifa Liang}, {and} \bibinfo{person}{Yun Yang}.} \bibinfo{year}{2022}\natexlab{}.
\newblock \showarticletitle{Pyramid: Enabling hierarchical neural networks with edge computing}. In \bibinfo{booktitle}{\emph{Proceedings of the ACM Web Conference 2022}}. \bibinfo{pages}{1860--1870}.
\newblock


\bibitem[HikRobot(2024)]%
        {hikcamera}
\bibfield{author}{\bibinfo{person}{HikRobot}.} \bibinfo{year}{2024}\natexlab{}.
\newblock \bibinfo{booktitle}{\emph{MV-CB060-10UM-S}}.
\newblock
\urldef\tempurl%
\url{https://www.hikrobotics.com/cn/machinevision/productdetail?id=3634}
\showURL{%
\tempurl}


\bibitem[Hinton(2015)]%
        {distill}
\bibfield{author}{\bibinfo{person}{Geoffrey Hinton}.} \bibinfo{year}{2015}\natexlab{}.
\newblock \showarticletitle{Distilling the Knowledge in a Neural Network}.
\newblock \bibinfo{journal}{\emph{arXiv preprint arXiv:1503.02531}} (\bibinfo{year}{2015}).
\newblock


\bibitem[Huang et~al\mbox{.}(2023)]%
        {huang2023re}
\bibfield{author}{\bibinfo{person}{Jin Huang}, \bibinfo{person}{Hui Guan}, {and} \bibinfo{person}{Deepak Ganesan}.} \bibinfo{year}{2023}\natexlab{}.
\newblock \showarticletitle{Re-thinking computation offload for efficient inference on IoT devices with duty-cycled radios}. In \bibinfo{booktitle}{\emph{Proceedings of the 29th Annual International Conference on Mobile Computing and Networking}}. \bibinfo{pages}{1--15}.
\newblock


\bibitem[Inc.(2023)]%
        {hvpm}
\bibfield{author}{\bibinfo{person}{Monsoon~Solutions Inc.}} \bibinfo{year}{2023}\natexlab{}.
\newblock \bibinfo{booktitle}{\emph{High Voltage Power Monitor}}.
\newblock
\urldef\tempurl%
\url{https://www.msoon.com/high-voltage-power-monitor}
\showURL{%
\tempurl}


\bibitem[Kan et~al\mbox{.}(2022)]%
        {kan2022improving}
\bibfield{author}{\bibinfo{person}{Nuowen Kan}, \bibinfo{person}{Yuankun Jiang}, \bibinfo{person}{Chenglin Li}, \bibinfo{person}{Wenrui Dai}, \bibinfo{person}{Junni Zou}, {and} \bibinfo{person}{Hongkai Xiong}.} \bibinfo{year}{2022}\natexlab{}.
\newblock \showarticletitle{Improving generalization for neural adaptive video streaming via meta reinforcement learning}. In \bibinfo{booktitle}{\emph{Proceedings of the 30th ACM International Conference on Multimedia}}. \bibinfo{pages}{3006--3016}.
\newblock


\bibitem[Kang et~al\mbox{.}(2017)]%
        {kang2017neurosurgeon}
\bibfield{author}{\bibinfo{person}{Yiping Kang}, \bibinfo{person}{Johann Hauswald}, \bibinfo{person}{Cao Gao}, \bibinfo{person}{Austin Rovinski}, \bibinfo{person}{Trevor Mudge}, \bibinfo{person}{Jason Mars}, {and} \bibinfo{person}{Lingjia Tang}.} \bibinfo{year}{2017}\natexlab{}.
\newblock \showarticletitle{Neurosurgeon: Collaborative intelligence between the cloud and mobile edge}.
\newblock \bibinfo{journal}{\emph{ACM SIGARCH Computer Architecture News}} \bibinfo{volume}{45}, \bibinfo{number}{1} (\bibinfo{year}{2017}), \bibinfo{pages}{615--629}.
\newblock


\bibitem[Kimmel(1999)]%
        {demosaic-0}
\bibfield{author}{\bibinfo{person}{R. Kimmel}.} \bibinfo{year}{1999}\natexlab{}.
\newblock \showarticletitle{Demosaicing: image reconstruction from color CCD samples}.
\newblock \bibinfo{journal}{\emph{IEEE Transactions on Image Processing}} \bibinfo{volume}{8}, \bibinfo{number}{9} (\bibinfo{year}{1999}), \bibinfo{pages}{1221--1228}.
\newblock
\href{https://doi.org/10.1109/83.784434}{doi:\nolinkurl{10.1109/83.784434}}


\bibitem[Kodak(1999)]%
        {kodak}
\bibfield{author}{\bibinfo{person}{Kodak}.} \bibinfo{year}{1999}\natexlab{}.
\newblock \bibinfo{booktitle}{\emph{Kodak Lossless True Color Image Suite}}.
\newblock
\urldef\tempurl%
\url{https://r0k.us/graphics/kodak/}
\showURL{%
\tempurl}


\bibitem[Kong et~al\mbox{.}(2023)]%
        {kong2023accumo}
\bibfield{author}{\bibinfo{person}{Z~Jonny Kong}, \bibinfo{person}{Qiang Xu}, \bibinfo{person}{Jiayi Meng}, {and} \bibinfo{person}{Y~Charlie Hu}.} \bibinfo{year}{2023}\natexlab{}.
\newblock \showarticletitle{AccuMO: Accuracy-centric multitask offloading in edge-assisted mobile augmented reality}. In \bibinfo{booktitle}{\emph{Proceedings of the 29th Annual International Conference on Mobile Computing and Networking}}. \bibinfo{pages}{1--16}.
\newblock


\bibitem[Lakshmi et~al\mbox{.}(2016)]%
        {lakshmi2016visually}
\bibfield{author}{\bibinfo{person}{M Lakshmi}, \bibinfo{person}{J Senthilkumar}, {and} \bibinfo{person}{Y Suresh}.} \bibinfo{year}{2016}\natexlab{}.
\newblock \showarticletitle{Visually lossless compression for Bayer color filter array using optimized vector quantization}.
\newblock \bibinfo{journal}{\emph{Applied Soft Computing}}  \bibinfo{volume}{46} (\bibinfo{year}{2016}), \bibinfo{pages}{1030--1042}.
\newblock


\bibitem[Li et~al\mbox{.}(2020)]%
        {li2020reducto}
\bibfield{author}{\bibinfo{person}{Yuanqi Li}, \bibinfo{person}{Arthi Padmanabhan}, \bibinfo{person}{Pengzhan Zhao}, \bibinfo{person}{Yufei Wang}, \bibinfo{person}{Guoqing~Harry Xu}, {and} \bibinfo{person}{Ravi Netravali}.} \bibinfo{year}{2020}\natexlab{}.
\newblock \showarticletitle{Reducto: On-camera filtering for resource-efficient real-time video analytics}. In \bibinfo{booktitle}{\emph{Proceedings of the Annual conference of the ACM Special Interest Group on Data Communication on the applications, technologies, architectures, and protocols for computer communication}}. \bibinfo{pages}{359--376}.
\newblock


\bibitem[Lin et~al\mbox{.}(2014)]%
        {lin2014microsoft}
\bibfield{author}{\bibinfo{person}{Tsung-Yi Lin}, \bibinfo{person}{Michael Maire}, \bibinfo{person}{Serge Belongie}, \bibinfo{person}{James Hays}, \bibinfo{person}{Pietro Perona}, \bibinfo{person}{Deva Ramanan}, \bibinfo{person}{Piotr Doll{\'a}r}, {and} \bibinfo{person}{C~Lawrence Zitnick}.} \bibinfo{year}{2014}\natexlab{}.
\newblock \showarticletitle{Microsoft coco: Common objects in context}. In \bibinfo{booktitle}{\emph{Computer Vision--ECCV 2014: 13th European Conference, Zurich, Switzerland, September 6-12, 2014, Proceedings, Part V 13}}. Springer, \bibinfo{pages}{740--755}.
\newblock


\bibitem[Liu et~al\mbox{.}(2019)]%
        {liu2019edge}
\bibfield{author}{\bibinfo{person}{Luyang Liu}, \bibinfo{person}{Hongyu Li}, {and} \bibinfo{person}{Marco Gruteser}.} \bibinfo{year}{2019}\natexlab{}.
\newblock \showarticletitle{Edge assisted real-time object detection for mobile augmented reality}. In \bibinfo{booktitle}{\emph{The 25th annual international conference on mobile computing and networking}}. \bibinfo{pages}{1--16}.
\newblock


\bibitem[Liu et~al\mbox{.}(2022a)]%
        {liu2022self}
\bibfield{author}{\bibinfo{person}{Shengzhong Liu}, \bibinfo{person}{Xinzhe Fu}, \bibinfo{person}{Maggie Wigness}, \bibinfo{person}{Philip David}, \bibinfo{person}{Shuochao Yao}, \bibinfo{person}{Lui Sha}, {and} \bibinfo{person}{Tarek Abdelzaher}.} \bibinfo{year}{2022}\natexlab{a}.
\newblock \showarticletitle{Self-cueing real-time attention scheduling in criticality-aware visual machine perception}. In \bibinfo{booktitle}{\emph{2022 IEEE 28th Real-Time and Embedded Technology and Applications Symposium (RTAS)}}. IEEE, \bibinfo{pages}{173--186}.
\newblock


\bibitem[Liu et~al\mbox{.}(2022b)]%
        {liu2022adamask}
\bibfield{author}{\bibinfo{person}{Shengzhong Liu}, \bibinfo{person}{Tianshi Wang}, \bibinfo{person}{Jinyang Li}, \bibinfo{person}{Dachun Sun}, \bibinfo{person}{Mani Srivastava}, {and} \bibinfo{person}{Tarek Abdelzaher}.} \bibinfo{year}{2022}\natexlab{b}.
\newblock \showarticletitle{Adamask: Enabling machine-centric video streaming with adaptive frame masking for dnn inference offloading}. In \bibinfo{booktitle}{\emph{Proceedings of the 30th ACM international conference on multimedia}}. \bibinfo{pages}{3035--3044}.
\newblock


\bibitem[Liu et~al\mbox{.}(2023)]%
        {liu2023sa}
\bibfield{author}{\bibinfo{person}{Xingyu Liu}, \bibinfo{person}{Pengfei Ren}, \bibinfo{person}{Yuchen Chen}, \bibinfo{person}{Cong Liu}, \bibinfo{person}{Jing Wang}, \bibinfo{person}{Haifeng Sun}, \bibinfo{person}{Qi Qi}, {and} \bibinfo{person}{Jingyu Wang}.} \bibinfo{year}{2023}\natexlab{}.
\newblock \showarticletitle{SA-Fusion: Multimodal Fusion Approach for Web-based Human-Computer Interaction in the Wild}. In \bibinfo{booktitle}{\emph{Proceedings of the ACM Web Conference 2023}}. \bibinfo{pages}{3883--3891}.
\newblock


\bibitem[Liu et~al\mbox{.}(2018)]%
        {liu2018deepn}
\bibfield{author}{\bibinfo{person}{Zihao Liu}, \bibinfo{person}{Tao Liu}, \bibinfo{person}{Wujie Wen}, \bibinfo{person}{Lei Jiang}, \bibinfo{person}{Jie Xu}, \bibinfo{person}{Yanzhi Wang}, {and} \bibinfo{person}{Gang Quan}.} \bibinfo{year}{2018}\natexlab{}.
\newblock \showarticletitle{DeepN-JPEG: A deep neural network favorable JPEG-based image compression framework}. In \bibinfo{booktitle}{\emph{Proceedings of the 55th annual design automation conference}}. \bibinfo{pages}{1--6}.
\newblock


\bibitem[Lu(2023)]%
        {lu2023object}
\bibfield{author}{\bibinfo{person}{Guoyu Lu}.} \bibinfo{year}{2023}\natexlab{}.
\newblock \showarticletitle{Object Detection Based on Raw Bayer Images}. In \bibinfo{booktitle}{\emph{2023 IEEE/RSJ International Conference on Intelligent Robots and Systems (IROS)}}. IEEE, \bibinfo{pages}{9582--9589}.
\newblock


\bibitem[Lu and Smith(2007)]%
        {lu2007augmented}
\bibfield{author}{\bibinfo{person}{Yuzhu Lu} {and} \bibinfo{person}{Shana Smith}.} \bibinfo{year}{2007}\natexlab{}.
\newblock \showarticletitle{Augmented reality e-commerce assistant system: trying while shopping}. In \bibinfo{booktitle}{\emph{Human-Computer Interaction. Interaction Platforms and Techniques: 12th International Conference, HCI International 2007, Beijing, China, July 22-27, 2007, Proceedings, Part II 12}}. Springer, \bibinfo{pages}{643--652}.
\newblock


\bibitem[Ma et~al\mbox{.}(2023)]%
        {ma2023leca}
\bibfield{author}{\bibinfo{person}{Tianrui Ma}, \bibinfo{person}{Adith~Jagadish Boloor}, \bibinfo{person}{Xiangxing Yang}, \bibinfo{person}{Weidong Cao}, \bibinfo{person}{Patrick Williams}, \bibinfo{person}{Nan Sun}, \bibinfo{person}{Ayan Chakrabarti}, {and} \bibinfo{person}{Xuan Zhang}.} \bibinfo{year}{2023}\natexlab{}.
\newblock \showarticletitle{Leca: In-sensor learned compressive acquisition for efficient machine vision on the edge}. In \bibinfo{booktitle}{\emph{Proceedings of the 50th Annual International Symposium on Computer Architecture}}. \bibinfo{pages}{1--14}.
\newblock


\bibitem[Mach and Becvar(2017)]%
        {offloadsurvey}
\bibfield{author}{\bibinfo{person}{Pavel Mach} {and} \bibinfo{person}{Zdenek Becvar}.} \bibinfo{year}{2017}\natexlab{}.
\newblock \showarticletitle{Mobile edge computing: A survey on architecture and computation offloading}.
\newblock \bibinfo{journal}{\emph{IEEE communications surveys \& tutorials}} \bibinfo{volume}{19}, \bibinfo{number}{3} (\bibinfo{year}{2017}), \bibinfo{pages}{1628--1656}.
\newblock


\bibitem[Menon et~al\mbox{.}(2006)]%
        {demosaic-2}
\bibfield{author}{\bibinfo{person}{Daniele Menon}, \bibinfo{person}{Stefano Andriani}, {and} \bibinfo{person}{Giancarlo Calvagno}.} \bibinfo{year}{2006}\natexlab{}.
\newblock \showarticletitle{Demosaicing with directional filtering and a posteriori decision}.
\newblock \bibinfo{journal}{\emph{IEEE Transactions on Image Processing}} \bibinfo{volume}{16}, \bibinfo{number}{1} (\bibinfo{year}{2006}), \bibinfo{pages}{132--141}.
\newblock


\bibitem[Mentzer et~al\mbox{.}(2020)]%
        {mentzer2020high}
\bibfield{author}{\bibinfo{person}{Fabian Mentzer}, \bibinfo{person}{George~D Toderici}, \bibinfo{person}{Michael Tschannen}, {and} \bibinfo{person}{Eirikur Agustsson}.} \bibinfo{year}{2020}\natexlab{}.
\newblock \showarticletitle{High-fidelity generative image compression}.
\newblock \bibinfo{journal}{\emph{Advances in Neural Information Processing Systems}}  \bibinfo{volume}{33} (\bibinfo{year}{2020}), \bibinfo{pages}{11913--11924}.
\newblock


\bibitem[Minnen and Singh(2020)]%
        {minnen2020channel}
\bibfield{author}{\bibinfo{person}{David Minnen} {and} \bibinfo{person}{Saurabh Singh}.} \bibinfo{year}{2020}\natexlab{}.
\newblock \showarticletitle{Channel-wise autoregressive entropy models for learned image compression}. In \bibinfo{booktitle}{\emph{2020 IEEE International Conference on Image Processing (ICIP)}}. IEEE, \bibinfo{pages}{3339--3343}.
\newblock


\bibitem[Muresan and Parks(2005)]%
        {demosaic-1}
\bibfield{author}{\bibinfo{person}{D~Darian Muresan} {and} \bibinfo{person}{Thomas~W Parks}.} \bibinfo{year}{2005}\natexlab{}.
\newblock \showarticletitle{Demosaicing using optimal recovery}.
\newblock \bibinfo{journal}{\emph{IEEE Transactions on Image Processing}} \bibinfo{volume}{14}, \bibinfo{number}{2} (\bibinfo{year}{2005}), \bibinfo{pages}{267--278}.
\newblock


\bibitem[Paszke et~al\mbox{.}(2019)]%
        {paszke2019pytorch}
\bibfield{author}{\bibinfo{person}{Adam Paszke}, \bibinfo{person}{Sam Gross}, \bibinfo{person}{Francisco Massa}, {and} \bibinfo{person}{Lerer et al.}} \bibinfo{year}{2019}\natexlab{}.
\newblock \showarticletitle{Pytorch: An imperative style, high-performance deep learning library}.
\newblock \bibinfo{journal}{\emph{Advances in neural information processing systems}}  \bibinfo{volume}{32} (\bibinfo{year}{2019}).
\newblock


\bibitem[Piekarski and Thomas(2002)]%
        {piekarski2002arquake}
\bibfield{author}{\bibinfo{person}{Wayne Piekarski} {and} \bibinfo{person}{Bruce Thomas}.} \bibinfo{year}{2002}\natexlab{}.
\newblock \showarticletitle{ARQuake: the outdoor augmented reality gaming system}.
\newblock \bibinfo{journal}{\emph{Commun. ACM}} \bibinfo{volume}{45}, \bibinfo{number}{1} (\bibinfo{year}{2002}), \bibinfo{pages}{36--38}.
\newblock


\bibitem[Redmon(2016)]%
        {yolo}
\bibfield{author}{\bibinfo{person}{J Redmon}.} \bibinfo{year}{2016}\natexlab{}.
\newblock \showarticletitle{You only look once: Unified, real-time object detection}. In \bibinfo{booktitle}{\emph{Proceedings of the IEEE conference on computer vision and pattern recognition}}.
\newblock


\bibitem[Ren et~al\mbox{.}(2021)]%
        {ren2021adaptive}
\bibfield{author}{\bibinfo{person}{Jie Ren}, \bibinfo{person}{Ling Gao}, \bibinfo{person}{Xiaoming Wang}, \bibinfo{person}{Miao Ma}, \bibinfo{person}{Guoyong Qiu}, \bibinfo{person}{Hai Wang}, \bibinfo{person}{Jie Zheng}, {and} \bibinfo{person}{Zheng Wang}.} \bibinfo{year}{2021}\natexlab{}.
\newblock \showarticletitle{Adaptive computation offloading for mobile augmented reality}.
\newblock \bibinfo{journal}{\emph{Proceedings of the ACM on Interactive, Mobile, Wearable and Ubiquitous Technologies}} \bibinfo{volume}{5}, \bibinfo{number}{4} (\bibinfo{year}{2021}), \bibinfo{pages}{1--30}.
\newblock


\bibitem[Richter et~al\mbox{.}(2021)]%
        {richter2021bayer}
\bibfield{author}{\bibinfo{person}{Thomas Richter}, \bibinfo{person}{Siegfried F{\"o}{\ss}el}, \bibinfo{person}{Antonin Descampe}, {and} \bibinfo{person}{Ga{\"e}l Rouvroy}.} \bibinfo{year}{2021}\natexlab{}.
\newblock \showarticletitle{Bayer CFA pattern compression with JPEG XS}.
\newblock \bibinfo{journal}{\emph{IEEE Transactions on Image Processing}}  \bibinfo{volume}{30} (\bibinfo{year}{2021}), \bibinfo{pages}{6557--6569}.
\newblock


\bibitem[Rush and Hubel(2003)]%
        {x3sensor}
\bibfield{author}{\bibinfo{person}{Allen Rush} {and} \bibinfo{person}{Paul Hubel}.} \bibinfo{year}{2003}\natexlab{}.
\newblock \showarticletitle{X3 sensor characteristics}.
\newblock \bibinfo{journal}{\emph{Journal of The Society of Photographic Science and Technology of Japan}} \bibinfo{volume}{66}, \bibinfo{number}{1} (\bibinfo{year}{2003}), \bibinfo{pages}{57--60}.
\newblock


\bibitem[Shenai et~al\mbox{.}(2011)]%
        {shenai2011virtual}
\bibfield{author}{\bibinfo{person}{Mahesh~B Shenai}, \bibinfo{person}{Marcus Dillavou}, \bibinfo{person}{Corey Shum}, \bibinfo{person}{Douglas Ross}, \bibinfo{person}{Richard~S Tubbs}, \bibinfo{person}{Alan Shih}, {and} \bibinfo{person}{Barton~L Guthrie}.} \bibinfo{year}{2011}\natexlab{}.
\newblock \showarticletitle{Virtual interactive presence and augmented reality (VIPAR) for remote surgical assistance}.
\newblock \bibinfo{journal}{\emph{Operative Neurosurgery}}  \bibinfo{volume}{68} (\bibinfo{year}{2011}), \bibinfo{pages}{ons200--ons207}.
\newblock


\bibitem[Tan and Le(2019)]%
        {tan2019efficientnet}
\bibfield{author}{\bibinfo{person}{Mingxing Tan} {and} \bibinfo{person}{Quoc Le}.} \bibinfo{year}{2019}\natexlab{}.
\newblock \showarticletitle{Efficientnet: Rethinking model scaling for convolutional neural networks}. In \bibinfo{booktitle}{\emph{International conference on machine learning}}. PMLR, \bibinfo{pages}{6105--6114}.
\newblock


\bibitem[Toderici et~al\mbox{.}(2015)]%
        {toderici2015variable}
\bibfield{author}{\bibinfo{person}{George Toderici}, \bibinfo{person}{Sean~M O'Malley}, \bibinfo{person}{Sung~Jin Hwang}, \bibinfo{person}{Damien Vincent}, \bibinfo{person}{David Minnen}, \bibinfo{person}{Shumeet Baluja}, \bibinfo{person}{Michele Covell}, {and} \bibinfo{person}{Rahul Sukthankar}.} \bibinfo{year}{2015}\natexlab{}.
\newblock \showarticletitle{Variable rate image compression with recurrent neural networks}.
\newblock \bibinfo{journal}{\emph{arXiv preprint arXiv:1511.06085}} (\bibinfo{year}{2015}).
\newblock


\bibitem[Toderici et~al\mbox{.}(2017)]%
        {toderici2017full}
\bibfield{author}{\bibinfo{person}{George Toderici}, \bibinfo{person}{Damien Vincent}, \bibinfo{person}{Nick Johnston}, \bibinfo{person}{Sung Jin~Hwang}, \bibinfo{person}{David Minnen}, \bibinfo{person}{Joel Shor}, {and} \bibinfo{person}{Michele Covell}.} \bibinfo{year}{2017}\natexlab{}.
\newblock \showarticletitle{Full resolution image compression with recurrent neural networks}. In \bibinfo{booktitle}{\emph{Proceedings of the IEEE conference on Computer Vision and Pattern Recognition}}. \bibinfo{pages}{5306--5314}.
\newblock


\bibitem[Wallace(1992)]%
        {wallace1992jpeg}
\bibfield{author}{\bibinfo{person}{Gregory~K Wallace}.} \bibinfo{year}{1992}\natexlab{}.
\newblock \showarticletitle{The JPEG still picture compression standard}.
\newblock \bibinfo{journal}{\emph{IEEE transactions on consumer electronics}} \bibinfo{volume}{38}, \bibinfo{number}{1} (\bibinfo{year}{1992}), \bibinfo{pages}{xviii--xxxiv}.
\newblock


\bibitem[Wang et~al\mbox{.}(2023a)]%
        {wang2023real}
\bibfield{author}{\bibinfo{person}{Hao Wang}, \bibinfo{person}{Hao Bao}, \bibinfo{person}{Liekang Zeng}, \bibinfo{person}{Ke Luo}, {and} \bibinfo{person}{Xu Chen}.} \bibinfo{year}{2023}\natexlab{a}.
\newblock \showarticletitle{Real-Time High-Resolution Pedestrian Detection in Crowded Scenes via Parallel Edge Offloading}. In \bibinfo{booktitle}{\emph{ICC 2023-IEEE International Conference on Communications}}. IEEE, \bibinfo{pages}{2173--2178}.
\newblock


\bibitem[Wang et~al\mbox{.}(2023b)]%
        {wang2023progressive}
\bibfield{author}{\bibinfo{person}{Ruiqi Wang}, \bibinfo{person}{Hanyang Liu}, \bibinfo{person}{Jiaming Qiu}, \bibinfo{person}{Moran Xu}, \bibinfo{person}{Roch Gu{\'e}rin}, {and} \bibinfo{person}{Chenyang Lu}.} \bibinfo{year}{2023}\natexlab{b}.
\newblock \showarticletitle{Progressive Neural Compression for Adaptive Image Offloading under Timing Constraints}. In \bibinfo{booktitle}{\emph{2023 IEEE Real-Time Systems Symposium (RTSS)}}. IEEE, \bibinfo{pages}{118--130}.
\newblock


\bibitem[Wootton(2005)]%
        {3ccd}
\bibfield{author}{\bibinfo{person}{Cliff Wootton}.} \bibinfo{year}{2005}\natexlab{}.
\newblock \bibinfo{booktitle}{\emph{A practical guide to video and audio compression: From sprockets and rasters to macro blocks}}.
\newblock \bibinfo{publisher}{Routledge}.
\newblock


\bibitem[Wu et~al\mbox{.}(2013)]%
        {wu2013current}
\bibfield{author}{\bibinfo{person}{Hsin-Kai Wu}, \bibinfo{person}{Silvia Wen-Yu Lee}, \bibinfo{person}{Hsin-Yi Chang}, {and} \bibinfo{person}{Jyh-Chong Liang}.} \bibinfo{year}{2013}\natexlab{}.
\newblock \showarticletitle{Current status, opportunities and challenges of augmented reality in education}.
\newblock \bibinfo{journal}{\emph{Computers \& education}}  \bibinfo{volume}{62} (\bibinfo{year}{2013}), \bibinfo{pages}{41--49}.
\newblock


\bibitem[Xiao et~al\mbox{.}(2021)]%
        {xiao2021task}
\bibfield{author}{\bibinfo{person}{Shuo Xiao}, \bibinfo{person}{Shengzhi Wang}, \bibinfo{person}{Zhenzhen Huang}, \bibinfo{person}{Tianyu Wang}, \bibinfo{person}{Wei Chen}, {and} \bibinfo{person}{Guopeng Zhang}.} \bibinfo{year}{2021}\natexlab{}.
\newblock \showarticletitle{Task offloading strategy of internet of vehicles based on stackelberg game}. In \bibinfo{booktitle}{\emph{Companion Proceedings of the Web Conference 2021}}. \bibinfo{pages}{52--56}.
\newblock


\bibitem[Xie and Kim(2019)]%
        {xie2019source}
\bibfield{author}{\bibinfo{person}{Xiufeng Xie} {and} \bibinfo{person}{Kyu-Han Kim}.} \bibinfo{year}{2019}\natexlab{}.
\newblock \showarticletitle{Source compression with bounded dnn perception loss for iot edge computer vision}. In \bibinfo{booktitle}{\emph{The 25th Annual International Conference on Mobile Computing and Networking}}. \bibinfo{pages}{1--16}.
\newblock


\bibitem[Yang et~al\mbox{.}(2023)]%
        {yang2023tangible}
\bibfield{author}{\bibinfo{person}{Simin Yang}, \bibinfo{person}{Ze Gao}, \bibinfo{person}{Reza Hadi~Mogavi}, \bibinfo{person}{Pan Hui}, {and} \bibinfo{person}{Tristan Braud}.} \bibinfo{year}{2023}\natexlab{}.
\newblock \showarticletitle{Tangible web: An interactive immersion virtual reality creativity system that travels across reality}. In \bibinfo{booktitle}{\emph{Proceedings of the ACM Web Conference 2023}}. \bibinfo{pages}{3915--3922}.
\newblock


\bibitem[Yao et~al\mbox{.}(2020)]%
        {yao2020deep}
\bibfield{author}{\bibinfo{person}{Shuochao Yao}, \bibinfo{person}{Jinyang Li}, \bibinfo{person}{Dongxin Liu}, \bibinfo{person}{Tianshi Wang}, \bibinfo{person}{Shengzhong Liu}, \bibinfo{person}{Huajie Shao}, {and} \bibinfo{person}{Tarek Abdelzaher}.} \bibinfo{year}{2020}\natexlab{}.
\newblock \showarticletitle{Deep compressive offloading: Speeding up neural network inference by trading edge computation for network latency}. In \bibinfo{booktitle}{\emph{Proceedings of the 18th conference on embedded networked sensor systems}}. \bibinfo{pages}{476--488}.
\newblock


\bibitem[Yim et~al\mbox{.}(2017)]%
        {yim2017augmented}
\bibfield{author}{\bibinfo{person}{Mark Yi-Cheon Yim}, \bibinfo{person}{Shu-Chuan Chu}, {and} \bibinfo{person}{Paul~L Sauer}.} \bibinfo{year}{2017}\natexlab{}.
\newblock \showarticletitle{Is augmented reality technology an effective tool for e-commerce? An interactivity and vividness perspective}.
\newblock \bibinfo{journal}{\emph{Journal of interactive marketing}} \bibinfo{volume}{39}, \bibinfo{number}{1} (\bibinfo{year}{2017}), \bibinfo{pages}{89--103}.
\newblock


\bibitem[Yuen et~al\mbox{.}(2011)]%
        {yuen2011augmented}
\bibfield{author}{\bibinfo{person}{Steve Chi-Yin Yuen}, \bibinfo{person}{Gallayanee Yaoyuneyong}, {and} \bibinfo{person}{Erik Johnson}.} \bibinfo{year}{2011}\natexlab{}.
\newblock \showarticletitle{Augmented reality: An overview and five directions for AR in education}.
\newblock \bibinfo{journal}{\emph{Journal of Educational Technology Development and Exchange (JETDE)}} \bibinfo{volume}{4}, \bibinfo{number}{1} (\bibinfo{year}{2011}), \bibinfo{pages}{11}.
\newblock


\bibitem[Zhang et~al\mbox{.}(2021)]%
        {zhang2021elf}
\bibfield{author}{\bibinfo{person}{Wuyang Zhang}, \bibinfo{person}{Zhezhi He}, \bibinfo{person}{Luyang Liu}, \bibinfo{person}{Zhenhua Jia}, \bibinfo{person}{Yunxin Liu}, \bibinfo{person}{Marco Gruteser}, \bibinfo{person}{Dipankar Raychaudhuri}, {and} \bibinfo{person}{Yanyong Zhang}.} \bibinfo{year}{2021}\natexlab{}.
\newblock \showarticletitle{Elf: accelerate high-resolution mobile deep vision with content-aware parallel offloading}. In \bibinfo{booktitle}{\emph{Proceedings of the 27th Annual International Conference on Mobile Computing and Networking}}. \bibinfo{pages}{201--214}.
\newblock


\end{thebibliography}


\appendix
\section*{Appendix}

\section{Sobel-Tenengrad Edge Clearness Value} 
\label{apdx:clearness}
The edge clearness value is calculated by the Sobel-tenengrad operator, which is commonly used to assess the sharpness or clarity of image edges. The Sobel operator is an edge detection method used to calculate the gradient of the image intensity in both the horizontal (x) and vertical (y) directions. The two convolution kernels (horizontal gradient $G_{x}$, vertical gradient $G_{y}$) for the Sobel operator are defined as follows:
\begin{equation}
    G_{x}=\begin{bmatrix}
        -1&0&1\\
        -2&0&2\\
        -1&0&1\\
    \end{bmatrix}
\end{equation}
\vspace{0.1cm}
\begin{equation}
    G_{y}=\begin{bmatrix}
        -1&-2&-1\\
        0&0&0\\
        1&2&1\\
    \end{bmatrix}
\end{equation}

The gradient magnitude at each pixel is then calculated using the following equation:
\begin{equation}
    G=\sqrt{G^{2}_{x}+G^{2}_{y}}
\end{equation}

This magnitude represents the edge strength at each pixel. Larger values indicate stronger edges, which typically correspond to sharper image features. Then we can calculate the Tenengrad focus measure $T$, which is used to quantify the overall sharpness of the image, by summing the squared gradient magnitudes across the entire image. Higher values of $T$ indicate a clear or sharper image, as they reflect stronger edges throughout the image. The formula is given by:
\begin{equation}
    T=\sum_{i,j}{G(i,j)^{2}}
\end{equation}

\section{Software Solution vs. Hardware Solution}
\label{apdx:softvshard}

The ISP hardware-based demosaic processing is the most commonly used solution. However, ISP hardware capable of processing high-resolution images either is costly (\ie Sony BIONZ XR) or does not have enough performance causing high latency. The sensor we use in ABO is widely used not only in smartphones but also in industrial scenes like assembler robots. According to our measurements, getting demosaiced frames straight from the ISP takes an extra 33ms, compared with using OpenCV.

We have also conducted additional experiments on this subject of hardware vs. software implementation. The image sensor we use supports native cropping (only reads pixels in a selected area). We set the read-out resolution to $2560\times1440$ and $1920\times1080$ (\ie 2K and 1080P, commonly chosen display options) for both RAW and RGB pipelines. It turns out that the ISP hardware demosaic still results in a 10.6-18.8 ms more latency compared with the OpenCV software demosaic, showing the insufficient high-resolution processing performance on low-cost camera modules.
\section{Detailed Structure of \model Neural Codec}
\label{apdx:modeulstructure}

\begin{figure*}[t!]
    \centering
    \includegraphics[width=1\textwidth]{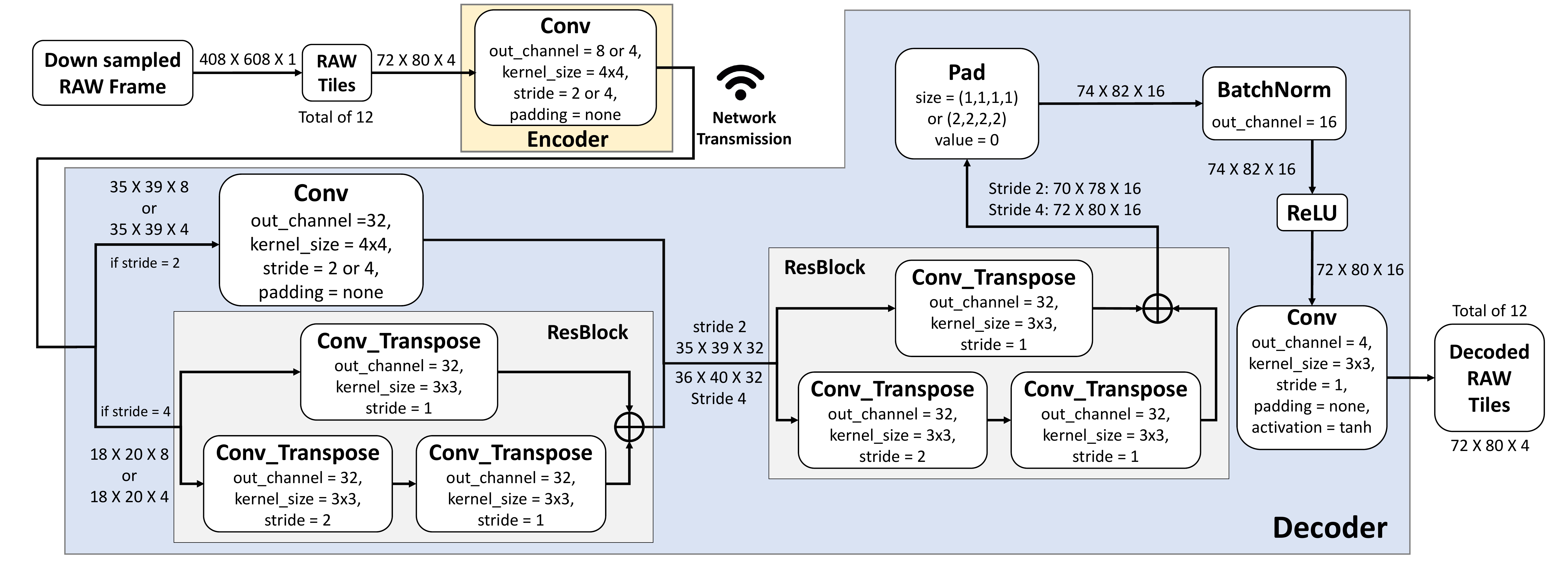}
    \vspace{-0.6cm}
    \caption{The configurable image neural codec in \model. }
    \label{fig:modelstruct}
\end{figure*}

The detailed model structure of \model neural codec is illustrated in Figure~\ref{fig:modelstruct}.
There are 4 configurations in total, decided by convolution kernel strides (2 and 4) and output channels (4 and 8) of the encoder. Upon being encoded and transmitted, the feature map will be divided into two types according to the stride and put into different decoder heads. The feature maps encoded with stride 2 will go through a single convolution layer, while the ones encoded with stride 4 will go through a ResBlock to be upscaled. At this point, the dimensions of the two types of feature maps are roughly aligned. They will go through two different padding layers to be exactly aligned on all dimensions after being output by the same ResBlock. After being further processed by other layers, the feature maps are finally decoded and ready to serve as input for the downstream DNN model.

\begin{figure*}[t!]
\centering
\begin{minipage}{.47\linewidth}
  \centering
    \includegraphics[width=0.9\linewidth]{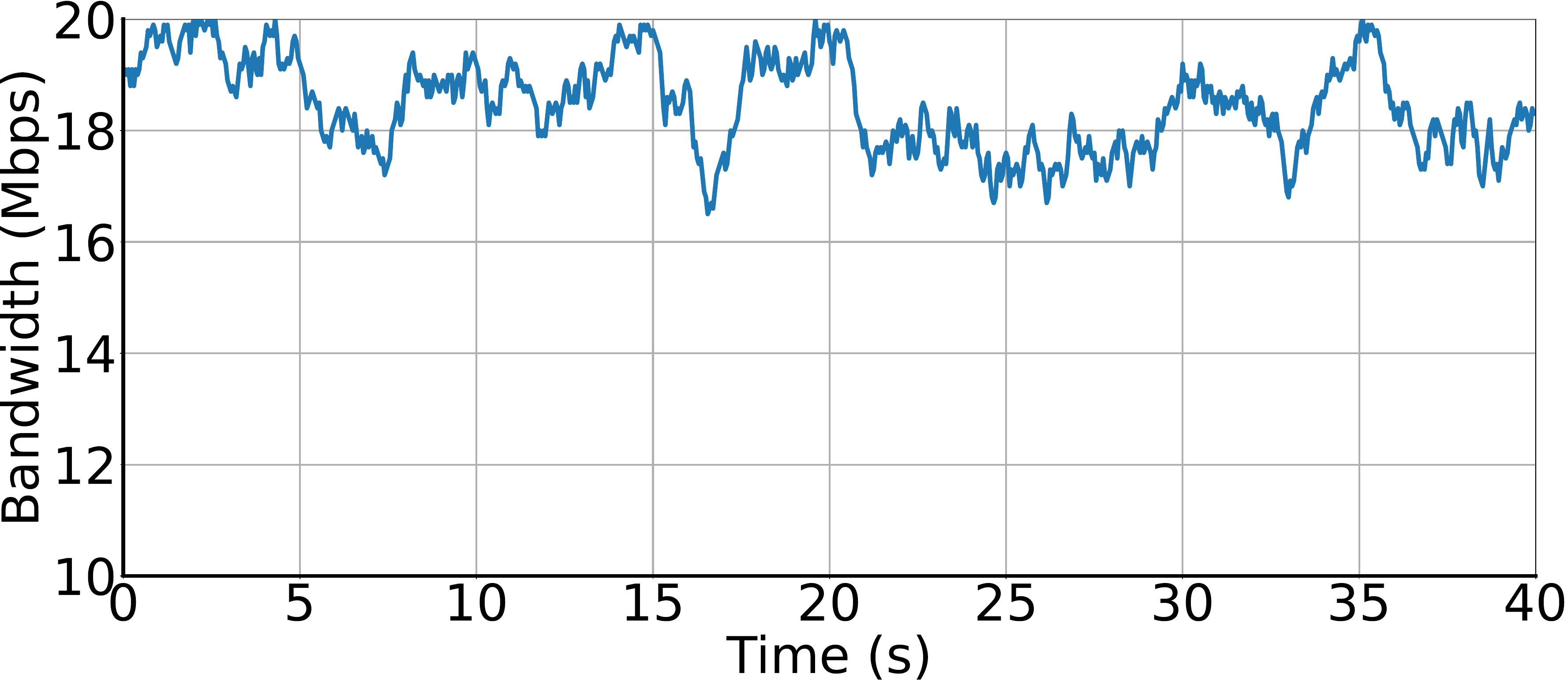}
    \vspace{-0.3cm}
    \caption{Random Generated Bandwidth Trace 1. }
    \Description{..}
    \label{fig:trace1}
\end{minipage}
\begin{minipage}{0.02\linewidth}

\end{minipage}
\begin{minipage}{.47\linewidth}
  \centering
    \includegraphics[width=0.9\linewidth]{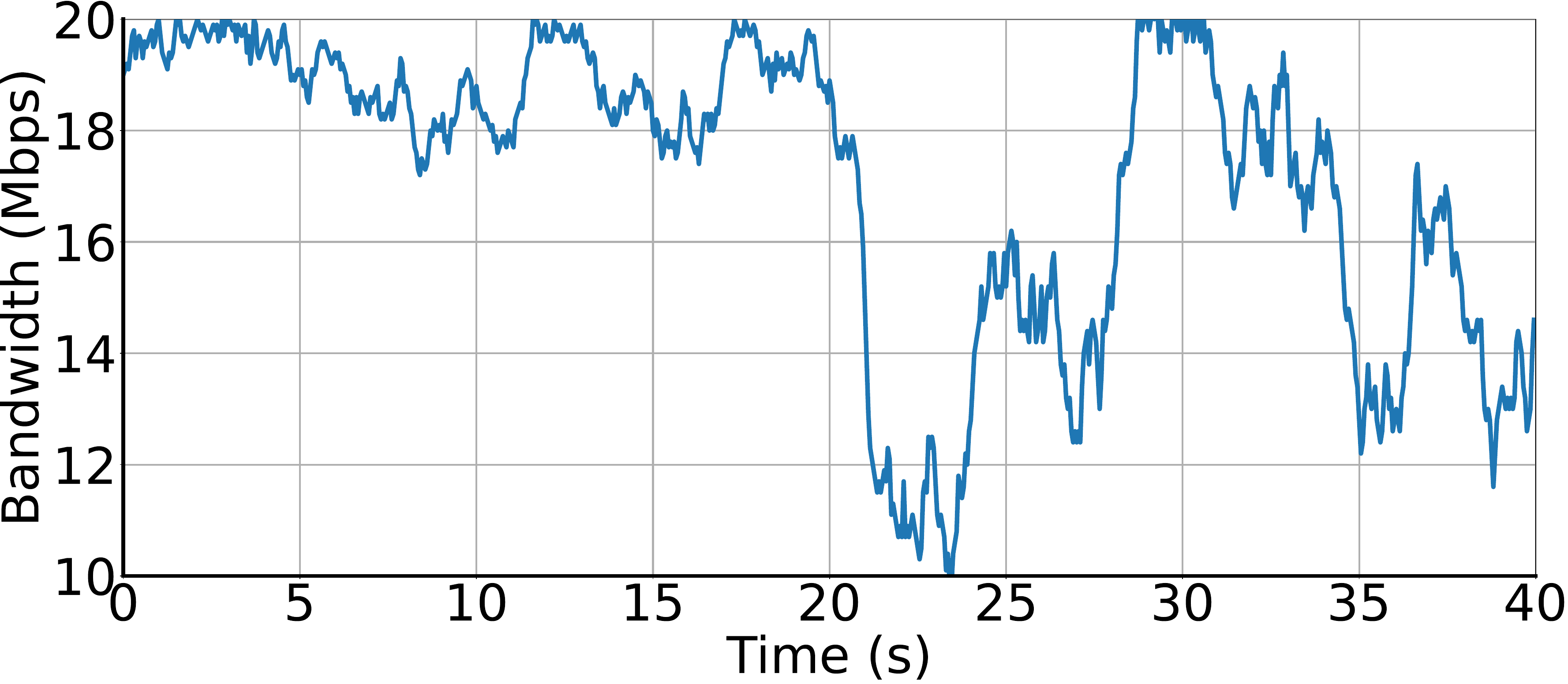}
    \vspace{-0.3cm}
    \caption{Random Generated Bandwidth Trace 2. }
    \Description{..}
    \label{fig:trace2}
\end{minipage}
\end{figure*}

\begin{figure*}[t!]
\centering
\begin{minipage}{.47\linewidth}
  \centering
    \vspace{0.2cm}
    \includegraphics[width=0.9\linewidth]{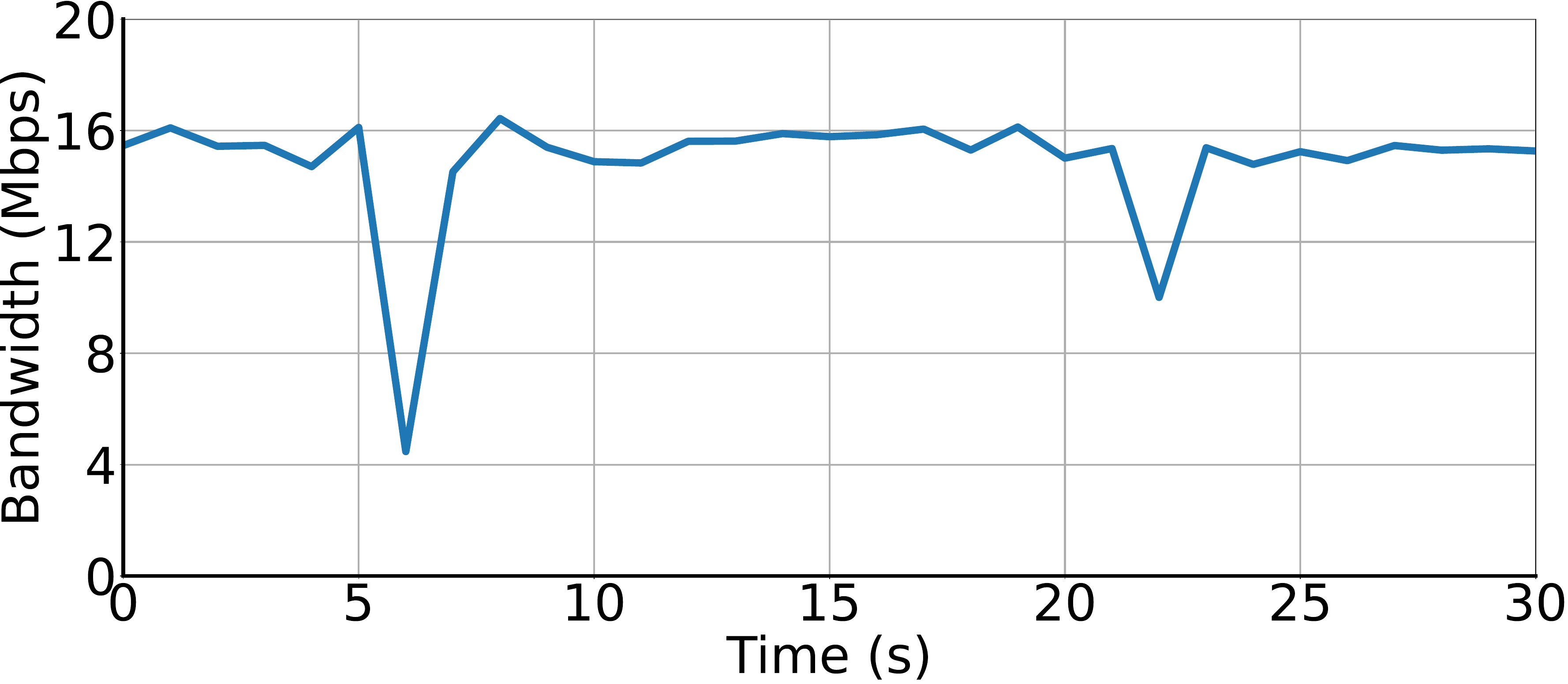}
    \vspace{-0.3cm}
    \caption{Real-World Recorded Bandwidth Trace 1. }
    \Description{..}
    \label{fig:trace3}
\end{minipage}
\begin{minipage}{0.02\linewidth}

\end{minipage}
\begin{minipage}{.47\linewidth}
  \centering
    \vspace{0.2cm}
    \includegraphics[width=0.9\linewidth]{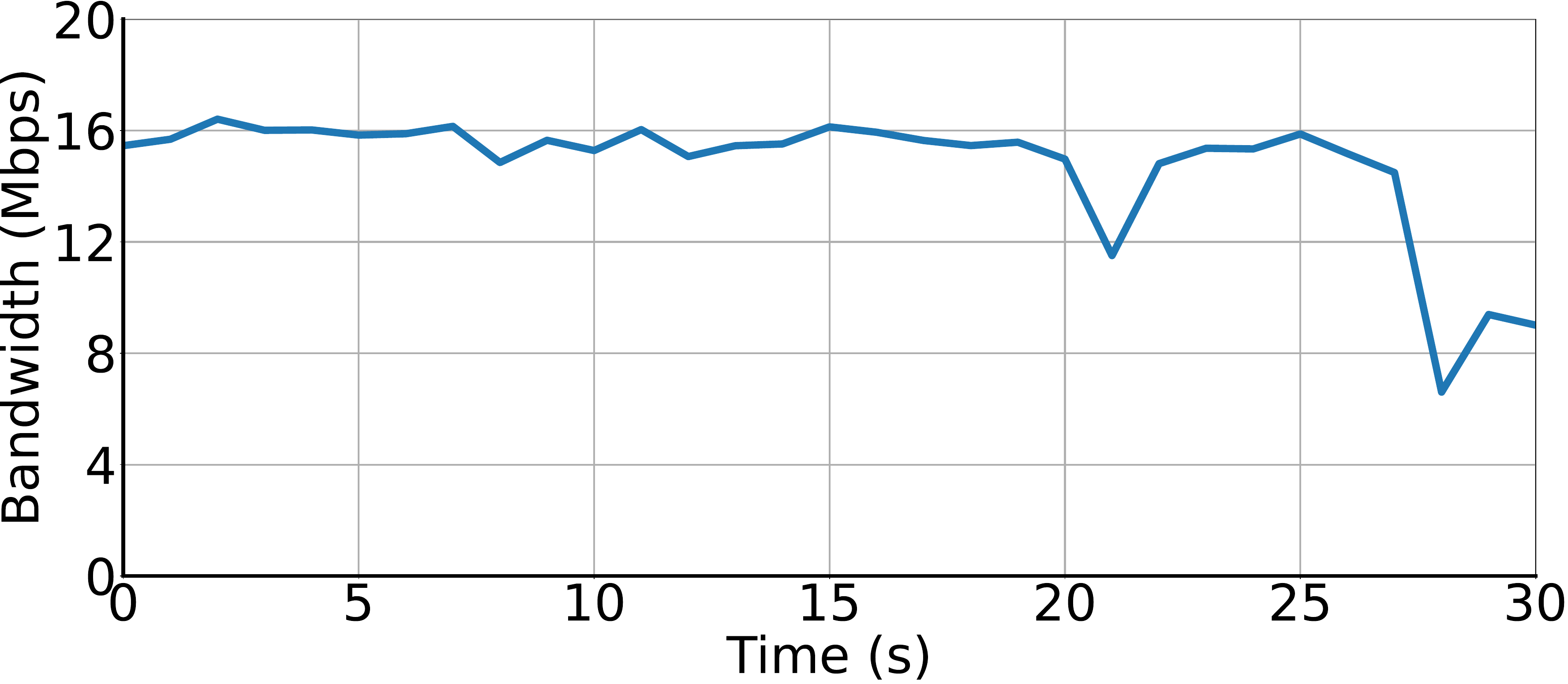}
    \vspace{-0.3cm}
    \caption{Real-World Recorded Bandwidth Trace 2. }
    \Description{..}
    \label{fig:trace4}
\end{minipage}
\end{figure*}

\section{Dataset Details}
\label{apdx:rawdataset}

In our experiments, the downstream vision task deployed on the server is YOLO-based object detection. Existing common datasets \cite{lin2014microsoft,deng2009imagenet} are all collected and labeled with RGB standards like JPEG, and simulating RAW images from such data is nearly impossible. On the other hand, existing RAW datasets \cite{abdorgchen2018learning, dong2022abandoning, kodak} either simply do not have the sufficient amount of images to create a persuading test environment, or are not designed for vision tasks but only compressive encoding. Thus, for experiments, we construct a new RAW dataset, the \model RAW dataset, with 17 types of common objects in campus offices, containing over 10,000 images with high-quality labeled bounding boxes. All the images are captured with a resolution of 3072*2048 under 8-bit non-packing Bayer mode with the digital gain set to zero to reduce any possible noises and secure image quality.
This dataset is designed not only to serve the object detection task but also to improve the robustness under common challenges in AR scenes like dim light or motion blur. To make the proxy of AR scenario more viable and improve the robustness, random movements of the camera and objects are introduced during the data collection process to improve the tolerance of motion blur in real-life AR scenarios. Roughly 5\% of the total dataset is collected under insufficient luminosity conditions to robustness under indoor environments. We also introduced several frames of pure black and white for gamma calibration.

However, the dataset is collected one frame at a time to maintain the variety of samples, meaning it cannot be used as frame streams since the frames are not consecutive. To evaluate \model, we also collected 5 segments of consecutive frames at a fixed 30 FPS frame rate using the prototype AR device. The segments are designed to include different possible real-world AR scenarios with various time lengths, simulating both short-burst and long-term services.

\section{Bandwidth Traces}
\label{apdx:bandtrace}

To evaluate \model under various network conditions, a binomial random algorithm generates half of the bandwidth traces, while the other half are recorded in real-world scenes~\cite{kan2022improving}. The traces are applied through a shell script using Linux Traffic Control (tc). The script is started along with the main Python script. Since the tc rules need to be erased before the next utilization, an apply-reset-apply cycle is longer than the time granularity of which when the trace is generated, meaning that the tc controlling script can only sample from the trace at the corresponding time. 
We visualize two of the randomly generated traces in Figure~\ref{fig:trace1} and~\ref{fig:trace2}, and two of the real-world recorded traces in Figure~\ref{fig:trace3} and~\ref{fig:trace4}.

\section{Details of Capability Experiments}
\label{apdx:capability}

\begin{figure}[t!]
    \centering
    \vspace{0.3cm}
    \includegraphics[width=\linewidth]{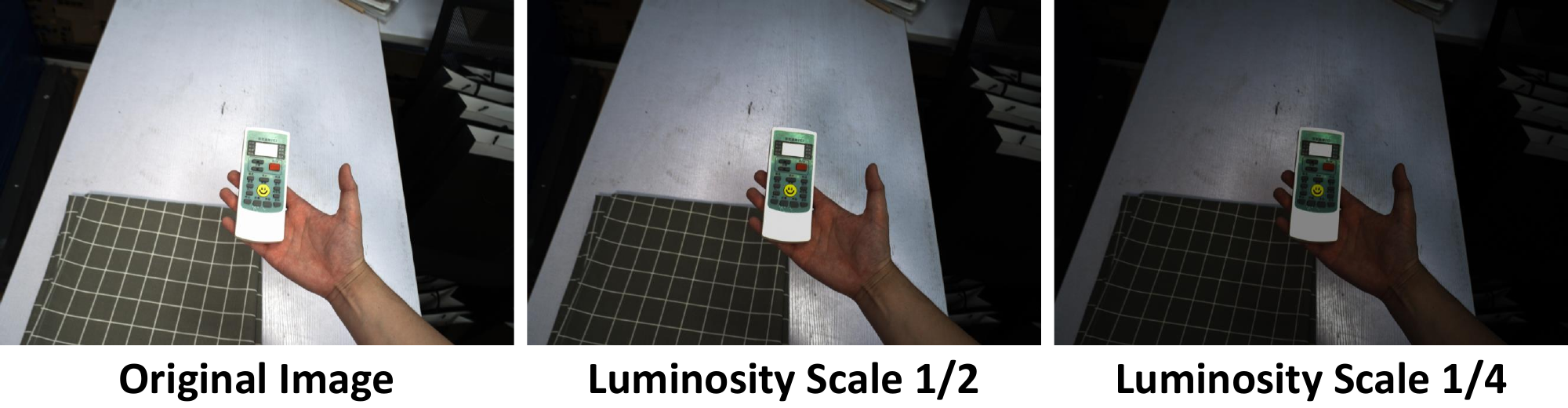}
    \vspace{-0.7cm}
    \caption{Example of Low-Light Frames}
    \label{fig:capability-lowlight}
\end{figure}

\begin{figure}[t!]
    \centering
    \vspace{0.1cm}
    \includegraphics[width=\linewidth]{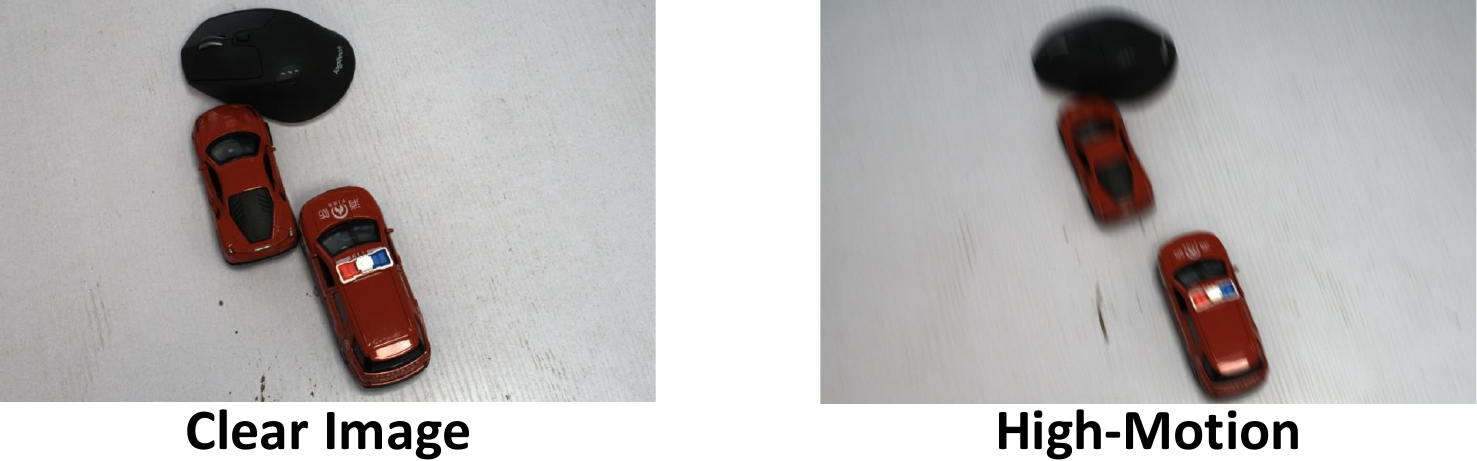}
    \vspace{-0.6cm}
    \caption{Example of High-Motion Frames}
    \label{fig:capability-motion}
\end{figure}

\begin{figure}[t!]
    \centering
    \vspace{0.2cm}
    \includegraphics[width=\linewidth]{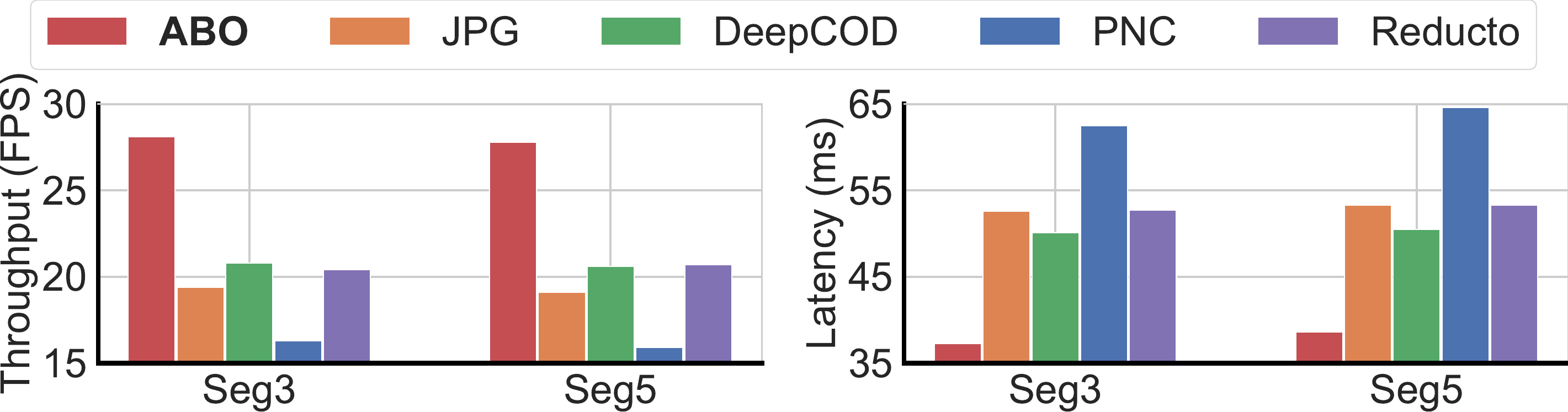}
    \vspace{-0.7cm}
    \caption{Efficiency under high-motion}
    \label{fig:motionblur-efficiency}
\end{figure}

Here we present the visualization of low-light frames (Figure~\ref{fig:capability-lowlight}) and high-motion ones (Figure~\ref{fig:capability-motion}). In Figure~\ref{fig:motionblur-efficiency}, we present the efficiency of \model under two high-motion test segments. 
The results show that under the pressure of high motion, \model not only achieves higher task accuracy but also consistently provides the highest throughput and the lowest latency.

\section{Limitations and Future Work} 
\label{apdx:discussion}
Here we briefly discuss the existing limitations we identified on \model that can be potential future directions for extension.

\textbf{Pretraining Overhead:} Though the neural codec can be universally utilized for all kinds of data, the downstream DNN model still needs to be trained with RAW frames of required tasks since the input channel has changed. There is also some room for improvement in \model. The possibility of tile-level codec configuration calibration remains, meaning that the bandwidth consumption can be further reduced. There is also the potential for using a scene-targeting online calibration module to further optimize the performance of the tile-selection module.

\textbf{Limited RAW Image Data:} The current \model design relies on large-scale RAW image datasets for pretraining. Although we have made a significant amount of effort to collect data simulating daily AR scenarios, its overall scale is still limited compared to standard image benchmarks like ImageNet~\cite{deng2009imagenet} and COCO~\cite{lin2014microsoft}. It drives us to think if we can use a large-scale RGB image dataset to pretrain the autoencoder (\ie extract the semantic feature patterns) and adapt it to RAW input with limited RAW images, so the data challenges can be greatly alleviated.

\textbf{Subframe Codec Calibration:} Current design of \model transmission controller only calibrates codec configuration on frame-level. There is a potential for using a scene-aware module to calibrate codec configuration on a tile level, thus the bandwidth consumption can be further reduced.

\textbf{Multi-Modal Fusion:} The great success of augmented reality not only comes from visual DNN analysis but also from close collaborations between multi-modal fusion (\eg image, audio, video) that creates an immersive experience. However, the current \model design only considers image data as the input but ignores the potential optimizations within multi-modal data streams. We believe the general philosophy of decoupling the edge offloading pipeline from any unnecessary preprocessing steps could fit into various data formats, with corresponding signal processing knowledge.

\textbf{Multi-Task Compatibility: } 
In practice, multiple DNN models can be simultaneously applied to analyze the offloaded frames, serving different applications. How to make sure their compatibility within the knowledge distillation of \model pretraining could be a challenge. Within this context, we believe a self-supervised learning paradigm (\eg contrastive learning or masked autoencoder) that seeks to learn general data semantics without task information could be a promising solution to enhance the generalizability of offloaded RAW frames to heterogeneous tasks.
\balance

\end{document}